\documentclass[prb,twocolumn,amsmath,amssymb]{revtex4}
\begin{document}
\author{Kevin Leung$^*$}
\affiliation{Surface and Interface Sciences Department, MS 1415}
\author{Susan B.~Rempe}
\affiliation{Nanobiology Department, MS 0895}
\author{O. Anatole von Lilienfeld}
\affiliation{Multiscale Dynamic Materials Modeling Department, MS 1322 \\
Sandia National Laboratories, Albuquerque, New Mexico 87185, USA}
\date{\today}
\title{{\it Ab initio} molecular dynamics calculations of ion hydration
free energies}

\input epsf
 
\begin{abstract}
 
We apply {\it ab initio} molecular dynamics (AIMD) methods in conjunction with
the thermodynamic integration (TI) or ``$\lambda$-path'' technique to compute
the intrinsic hydration free energies of Li$^+$, Cl$^-$, and Ag$^+$
ions.  Using the Perdew-Burke-Ernzerhof (PBE) functional, adapting methods
developed for classical force field applications, and with consistent
assumptions about surface potential ($\phi$) contributions, we obtain
absolute AIMD hydration free energies ($\Delta G_{\rm hyd}$) within a few
kcal/mol, or better than 4\%, of Tissandier {\it et al.}'s
[J.~Phys~Chem.~A {\bf 102}, 7787 (1998)]
experimental values augmented with the SPC/E water model $\phi$ predictions.
The sums of Li$^+$/Cl$^-$ and Ag$^+$/Cl$^-$ AIMD $\Delta G_{\rm hyd}$, which
are not affected by surface potentials, are within 2.6\% and 1.2~\% of
experimental values, respectively.  We also report the free energy changes
associated with the transition
metal ion redox reaction Ag$^+$ + Ni$^{+}$ $\rightarrow$ Ag + Ni$^{2+}$ in
water.  The predictions for this reaction suggest that existing
estimates of $\Delta G_{\rm hyd}$ for unstable radiolysis intermediates
such as Ni$^{+}$ may need to be extensively revised.
\vspace{0.2in}\\
$^*$email: {\tt kleung@sandia.gov}

\end{abstract}
 
\maketitle

\section{Introduction}
 
Accurate predictions of hydration free energies of ions and molecules are
crucial for modeling chemical and biochemical reactions in water and the
adsorption of ionic species at water-material interfaces and inside
nanopores.\cite{intro}  
State-of-the-art Density Functional Theory (DFT)-based {\it ab initio}
molecular dynamics (AIMD) simulations allow modeling the breaking and
making of chemical bonds, as well as molecular polarizability.  Direct use
of AIMD to predict ion hydration free energies, $\Delta G_{\rm hyd}$,
will have significant impact on computational electrochemistry,
biophysics, desalination, energy storage applications, corrosion
studies, and geochemistry.  AIMD simulations have already
been extensively applied to study the hydration structure of
ions,\cite{structure,li_rempe,sabo,varma} in many cases leading to more
accurate predictions of the hydration number than classical force field
methods.  At the same time, using hydration structure information
plus DFT and quantum chemistry calculations, the quasi-chemical method
has been applied to predict highly accurate $\Delta G_{\rm hyd}$ for ions in
water and biological binding sites.\cite{quasichem} In this manuscript,
we generalize and apply $\Delta G_{\rm hyd}$ methods developed for classical
force fields to AIMD simulations.  In some cases, our work can be related
to ``alchemical'' potentials within the context of molecular grand-canonical
ensemble DFT that allows variations of atomic numbers and electron
numbers.\cite{anatole1}  
 
Many of the techniques we use for predicting AIMD
$\Delta G_{\rm hyd}$ have non-DFT precedents.
In classical force field treatments of hydrated ions, $\Delta G_{\rm hyd}$
at infinite ion dilution has been
successfully computed\cite{hummer_mono,na8,garde,classical}
using the thermodynamic integration (TI) method,\cite{ti0,ti} 
\begin{equation}
\Delta G_{\rm hyd}= \int_0^1 d\lambda \langle \frac{d H(\lambda)}{d
\lambda} \rangle_\lambda \hspace*{0.2in},  \label {eq1}
\end{equation}
or free energy perturbation\cite{fep} and closely related techniques.
Here $0 \leq \lambda \leq 1$ interpolates between the initial and final
systems, $H(\lambda)$ is the Hamiltonian as $\lambda$ varies, the brackets
denote equilibrium sampling with the Boltzmann factor
$\exp[-\beta H(\lambda)]$, and $\beta=1/k_{\rm B}T$.  For obvious reasons,
the method is also called ``$\lambda$-path integration.''\cite{anatole1}  
$\Delta G_{\rm hyd}$ is a state property, independent of the interpolation
pathway.  Force field parameters for ions are generally fitted with
a specific water model (e.g., SPC/E\cite{spce}) to reproduce
experimental $\Delta G_{\rm hyd}$ values.  
In simulations of monoatomic ions M with charge $q$, $\lambda$ is
conveniently set to be proportional to $q$ in Eq.~\ref{eq1}
such that the ion is ``charged up'' linearly from M$^{0}$ to M$^{q+}$.

Two critical theoretical advances have enabled direct comparisons
of predicted $\Delta G_{\rm hyd}$ with tabulated data.
(A) The long-range nature of coulomb
interactions means a significant simulation cell size dependence arises 
when using Ewald summations.\cite{note1} This dependence derives
from the interactions of
an ion with its images as well as with the neutralizing background in a
charged simulation cell.  To remove this dependence, Hummer, Pratt, and
Garcia devised a monopole correction so effective that even an 8-water
simulation cell containing a Na$^+$ ion already yields $\Delta G_{\rm hyd}$
well converged with system size.\cite{hummer_mono,na8}  (B) Comparison with
experiments effectively entails bringing an ion from vacuum at infinity into
the bulk liquid water region.  A surface potential, $\phi$, materializes at
the liquid-vapor interface, leading to a shift in the ion free energy $q\phi$
in the aqueous phase.\cite{pratt_sur,tildes,marsman}
Accounting for the surface potential, the calculated absolute ion hydration
free energy, which may not be measurable,\cite{pratt_sur} becomes
\begin{equation}
\Delta G_{\rm tot} = \Delta G_{\rm Ewald} + q (\phi_d + \phi_q). \label{tot}
\end{equation}
Here $\Delta G_{\rm Ewald}$ is the hydration free energy computed using
standard Ewald summation which assumes a zero average electrostatic
potential inside the simulation cell.\cite{saunders}  $\phi_d$
and $\phi_q$ are the dipolar and quadrupolar (or ``spherical second
moment'') contributions to the surface potential $\phi$.  
Some reported experimental data have subtracted the effect
of this potential\cite{marcus} while others have not.\cite{tiss}

The rapid convergence of $\Delta G_{\rm hyd}$ with simulation cell size 
(A) significantly facilitates the application of this $\Delta G_{\rm hyd}$
formalism to computationally costly DFT-based AIMD simulations.
Special attention should be paid to the surface potential contribution (B)
in AIMD settings.
Unlike classical models for water, $\phi=\phi_d+\phi_q$ has not yet been
predicted for AIMD water (e.g., computed with a generalized-gradient
approximated (GGA) Kohn-Sham functional such as Perdew-Burke-Ernzerhof
(PBE)\cite{pbe}).  Such a calculation would entail a large
simulation cell depicting the interface and long sampling trajectories.
Furthermore, as the liquid water density affects
$\phi_q$,\cite{pratt_sur,tildes,dang,marsman} the effectiveness of such
a calculation may further be limited by the fact that bulk GGA
water may not exhibit 1.0 g/cm$^3$ density.\cite{mundy,siepmann}
Although $\phi_d$ and $\phi_q$ are not independent --- they require
a common choice of molecular center, typically taken to be the oxygen
atom of water molecules --- the quantity $\phi_q$ has recently
been computed for PBE water using maximally localized Wannier
functions.\cite{marsman}  This piece of information is important for DFT-based
calculations because $\Delta G_{\rm Ewald}$ itself is an ambiguous
quantity whose value depends on whether the pseudopotential contains
core electrons, while $\Delta G_{\rm Ewald} + q\phi_q$ is independent of
such DFT details.  We therefore redefine
\begin{equation}
\Delta G_{\rm hyd} = \Delta G_{\rm Ewald} + q \phi_q. \label{hyd}
\end{equation}


To further enable comparison with experimental data in
Ref.~\onlinecite{marcus}, which contain no surface potential
contributions, we add $q\phi_q$=-19.7$q$~kcal/mol, the quadrupole
moment value for SPC/E water at 1.00~g/cc density when the
oxygen site is chosen as the molecular center.  This is
appropriate because $\Delta G_{\rm hyd}$ for various ions
have been fitted to Ref.~\onlinecite{marcus} using
the SPC/E water model\cite{garde} or the very similar SPC
model.\cite{hummer_mono,na8}  In effect, we are comparing
AIMD $\Delta G_{\rm hyd}$ with {\it SPC/E calculations
fitted to the data of Ref.~\onlinecite{marcus}.}
For the data tabulated in Ref.~\onlinecite{tiss}, which
contain the surface potential term $q (\phi_d + \phi_q)$, we subtract
$q\phi_d=4.8 q$~kcal/mol estimated using SPC/E water model-based
water-vapor interface molecular dynamics calculations.\cite{tildes}
Although an investigation of $\phi_d$ predicted with different
methods is not the focus of this work, accurate DFT methods and
accurate force fields should yield similar, reliable $\phi_d$.
Even if there exists a 50\% uncertainty in
this SPC/E $\phi_d$ estimate, $\Delta G_{\rm hyd} + q\phi_d$ in water will
be affected by only $\sim 2.4 |q|$~kcal/mol.  
Indeed, the much used SPC and the TIP4P water models yield
$\phi_d$=5.5 and 7.1~kcal/mol/$|e|$, respectively,\cite{barr,brod,note9}
which are slightly different from the SPC/E $\phi_d$.
The discrepancies among these models can be taken as a measure of
the systematic uncertainty associated with our $\phi_d$ assignments.

Finally, experimental data for moving ions from vacuum into
aqueous solution are referenced to their respective standard states,
i.e., gas phase ions at 1.0~atm.~pressure and hydrated ions
at 1.0~M concentration.  To be consistent with the infinite dilution
limit $\Delta G_{\rm hyd}$ predicted in this work, $C^{(0)}=1.9$~kcal/mol is
further subtracted from tabulated $\Delta G_{\rm Tiss}$ for all ions
regardless of their charges to account for the volume change included
in the experimental data.  Due to a sign problem,\cite{quasichem}
$2 C^{(0)}$~kcal/mol needs to be subtracted from
$\Delta G_{\rm Marcus}$ for this purpose.



To summarize, we compare our AIMD $\Delta G_{\rm hyd}$ (Eq.~\ref{hyd}) with
$\Delta G_{\rm Marcus} + q\phi_q^{\rm SPC/E}-2C^{(0)}$~kcal/mol and 
$\Delta G_{\rm Tiss} - q\phi_d^{\rm SPC/E}-C^{(0)}$~kcal/mol,
where $\Delta G_{\rm Marcus}$ and $\Delta G_{\rm Tiss}$ are the values
listed in Refs.~\onlinecite{marcus} and~\onlinecite{tiss}, respectively.

Note that the proton is often used as a reference for hydration free
energies.\cite{truhlar}  Referencing the predicted $\Delta G_{\rm hyd}$ 
of ions with that of H$^+$ computed in the same way circumvents the
need to estimate $\phi$.  In AIMD settings, however, an excess proton can
migrate from one H$_2$O to another.  Therefore we have not yet attempted to
compute this proton $\Delta G_{\rm hyd}$.

For test cases, we consider Li$^+$ and Cl$^-$.  The Li$^+$ ion hydration
structure and hydration free energies have been extensively studied using
AIMD and quasi-chemical methods, respectively.\cite{li_rempe}  Computing
the $\Delta G_{\rm hyd}$ of Cl$^-$ further allows us to predict the summed
$\Delta G_{\rm hyd}$ of the monovalent Li$^+$/Cl$^-$ pair,
where the surface potential terms cancel and the result 
contains less systematic uncertainty.  We show that this summed value is
at worst within 2.6\% of experimental results.\cite{marcus,tiss}

We also study the change in hydration free energies associated with 
\begin{equation}
{\rm Ag}^+ + {\rm Ni}^{+} \rightarrow {\rm Ag} + {\rm Ni}^{2+}, \label{agni}
\end{equation}
and the corresponding electrochemical half-cell reactions,
\begin{eqnarray}
{\rm Ag} &\rightarrow& {\rm Ag}^+ ; \hspace*{0.2in} {\rm and} \label{ag} \\
{\rm Ni}^{+} &\rightarrow& {\rm Ni}^{2+} . \label{ni}
\end{eqnarray}
These reactions are pertinent not only to elementary electrochemical
processes, but also to the initial stages of
nano alloy synthesis by radiolysis.\cite{nenoff,nanosyn}
$\gamma$ irradiation of mixed electrolytic aqueous solutions releases
secondary electrons that reduce the metal ions to atoms or lower oxidation
state ions. These reduced species readily coalesce to form clusters.  In the
case of a mixed Ag(I)/Ni(II) solution, the exothermicity of Eq.~\ref{agni}
will determine whether reduced Ni species are readily re-oxidized by Ag$^+$
in the solution --- a side reaction that hinders nano-alloy cluster formation.
AIMD is an attractive route to estimate the redox free energies associated with
Ni(I) species, which exhibit short lifetimes and are difficult to probe
experimentally.

Apart from the ability to compare AIMD $\Delta G_{\rm hyd}$ with
quasi-chemical theory\cite{quasichem,asthagiri} and potentially 
extend DFT-based absolute hydration free energy calculations 
to inhomogeneous media, this work is important due to its 
close relationship to recent
theoretical advances.  One is the alchemical $\lambda$-path integration
technique recently formulated within a DFT/AIMD-based molecular grand
canonical ensemble scheme,\cite{anatole1} which accounts for changes
in pseudopotentials as well as the number of electrons.
As long as the pseudopotential replaces all core electrons in the ion,
$\Delta G_{\rm hyd}$ TI calculations are very similar within AIMD and the SPC/E
model treatments of water.  More complex treatments are required, however,  when
ion insertion into the solvent involves not only changes in the ionic
pseudopotential, but also injection of electrons.\cite{anatole1}
This alchemical path technique has been applied to quantum mechanics/molecular
mechanics (QM/MM) simulations of
electron transfer reactions of aqueous metal complexes
(Fe(II/III) and Ru(II/III)).\cite{zeng}  Our work is even more closely related
to purely AIMD-based computational electrochemistry.\cite{seealso}
Here the electron transfer processes are similar to those in
Ref.~\onlinecite{zeng}, but all water molecules are treated with DFT methods,
and the long-range electrostatics are fundamentally different from those
in QM/MM calculations.  Our computational approach treats the ionization
potential and the ion hydration free energy contributions to the redox
potential separately.  While it is
based on and derives its rigor from theories well established with classical
force field hydration treatments (e.g., Eq.~\ref{hyd}), our thermodynamic
method has not been extended to estimate the fluctuating gaps that are
necessary for calculating reaction rates via the Marcus theory.\cite{seealso}
 
\section{Method}
 
\subsection{{\tt VASP} calculations}
 
We apply the Vienna atomistic simulation package ({\tt VASP})\cite{vasp} version
4.6 with a modified {\tt pot.F},\cite{marsman} the PBE
exchange correlation functional,\cite{pbe} projected-augmented wave
(PAW) pseudopotentials\cite{paw0,paw} (PP) with only valence electrons
for Li, Cl, H, and O atoms, and Ag and Ni PPs that include pseudovalent
4p and 3p electrons.
Two protocols to generate {\tt VASP} AIMD trajectories for Li$^+$
solvated in water are applied.  For the ion plus 32-water simulations, we
use a cell size of 9.855~\AA\, corresponding to a water density of 1.0~g/cc,
a 0.25~fs time step, an energy cutoff of 400~eV, and
a Born-Oppenheimer convergence of 10$^{-6}$~eV at each time step.  
For 64-water simulations, the corresponding parameters are 12.417~\AA\,
(1.0~g/cc), 0.5~fs, 500~eV, and 10$^{-7}$ eV, respectively.  
These settings limit the temperature drifts to 1~and~0.5~K/ps,
respectively.  The trajectory length for each value of $q$ is at least
40~ps in 2-point TI calculations and at least 30~ps for 6-point TI.
Initial configurations are pre-equilibrated using the
SPC/E water model and ion force fields\cite{garde} with charges scaled
to the net charge of the corresponding AIMD simulation cells.  A Nose
thermostat is applied, setting T=400~K, which is needed for the PBE functional
to describe experimental liquid water at room temperature.\cite{water400}
The deuterium mass is adopted for all protons to allow a larger time step,
although the H mass is assumed whenever water density is reported.
Ag$^+$ and Ni$^{2+}$ simulations are performed at 0.99 g/cc water
density while the Cl$^-$ simulation is at 1.0 g/cm$^3$
density; these simulation cells all contain 32 H$_2$O
molecules, and the time step, energy cutoff, and convergence
criteria used are analogous to those for Li$^+$/32 H$_2$O.

\subsection{Visualizing Electronic Isosurfaces}

\label{anatole}

Electronic isosurfaces and integrated changes in electron density,
$\Delta(x) = \int dy dz [\rho(x,y,z)_{n} - \rho(x,y,z)_c]$
as functions of spatial coordinate $x$, are also computed and depicted
for Li$^{q+}$ in water for various values of $q$.
The depicted geometries are snapshots taken at the
end of the 32-water PBE simulations.  These results
are obtained using the code {\tt CPMD},\cite{cpmd}
the PBE functional,\cite{pbe} pseudopotentials from Ref.~\onlinecite{SGPPKrack},
and a cutoff of 100 Ry (1361~eV).
$\rho_c$ refers to the electron density obtained by
minimizing the energy within the indicated charge.  As with {\tt VASP},
{\tt CPMD} uses an opposite background charge to neutralize the system within
the periodically replicated simulation cells.
$\rho_n$ corresponds to the density of the same geometry but
with the charged species replaced by a neutral He atom.

\subsection{Li$^+$ thermodynamic integration}
To implement Eq.~\ref{eq1} for Li$^+$, we generate integrand values at
different $q$ values according to two different integration formulas: a
two-point Gaussian quadrature and a six-point trapezoidal rule.  To that end,
AIMD trajectories apply a Li$^+$ pseudopotential (which contains no core
electrons) globally scaled by Gaussian quadrature values $q$=0.211325 and
0.788675.  This procedure 
is analogous to the scaling of the ionic charges in classical force field
molecular dynamics calculations of hydration free energies.\cite{asthagiri}
In addition, $q$=0.1, 0.4, 0.6, and 1.0 are considered.  Using these
6 points, a cubic least-squared fit is applied to extrapolate the integrand
value to $q$=0.\cite{note3}  These steps yield 6 almost evenly spaced
integration points needed to implement a trapezoidal rule integration.

\begin{figure}
\centerline{\hbox{\epsfxsize=3.00in \epsfbox{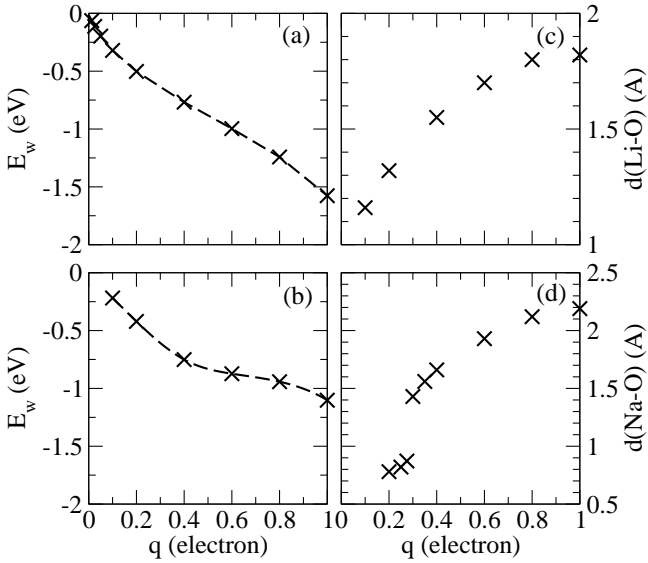}}}
\caption[]
{\label{fig1} \noindent
The binding energies and optimized distances between a H$_2$O molecule and
{\tt VASP} PBE pseudopotentials globally scaled by a factor of $0 < q \leq 1$.  
\hbox{(a) \& (c)}: Li$^+$; \hbox{(b)\&(d)}: Na$^+$.  The pseudopotentials
have no core electrons.  Dashed lines are cubic spline fits.
Na$^+$ is meant as a counter example to Li$^+$ for gas
phase behavior; its behavior in water will not be
the focus of this work.
}
\end{figure}

Figure~\ref{fig1}a shows that the scaled {\tt VASP} Li$^+$ pseudopotential 
behaves to some extent like a classical force field Li$^{q+}$; its
binding energy with one H$_2$O molecule scales roughly linearly with $q$
except at very small $q$.  The optimal Li-O$_{\rm water}$ distance also shrinks 
smoothly
with decreasing $q$ (Fig.~\ref{fig1}c).  In contrast, Fig.~\ref{fig1}b
shows that the scaled {\tt VASP} PBE Na$^+$ exhibits water binding energies
that deviate more strongly from linearity.  Furthermore, the optimal
$q$-scaled Na$^{+}$-OH$_2$ distance sharply decreases to 0.87~\AA\, at
$q \approx 0.29$, which suggests the formation of an anomalous covalent
bond beyond $q < 0.29$ (Fig.~\ref{fig1}d).  For efficient AIMD
$\Delta G_{\rm hyd}$ simulations, a pathway should be chosen such that
at the selected simulation points, electron transfer or unphysical
chemical bonding between the scaled pseudopotential and H$_2$O is avoided.
 
The AIMD trajectory is sampled every 0.1~ps.  At such intervals, we use
a finite difference method to compute $d H(q)/d q 
=[H(q+\Delta q/2)-H(q-\Delta q/2)]/q $ at fixed atomic
configurations.  Here $H(q)$ is the total potential energy of the simulation
cell predicted using {\tt VASP}.  When taking finite derivatives, $\Delta q$ values
of 0.025 and 0.050 yield Li$^+$ hydration free energies that agree to within
0.5~kcal/mol.  Evaluating $\langle d H(q)/dq \rangle_q$ using
400~eV and 500~eV cutoffs lead to indistinguishable results.
 
The derivative is corrected for finite size effects by adding the
Ewald correction to the energy, $\alpha q^2 /2 L$ at each $q$, where $\alpha$
is the Madelung constant, to the Li$^+$-plus-water {\tt VASP} energies
(issue ``(A)'' discussed in the introduction\cite{hummer_mono}).
The quadrupole moment correction $q\phi_q$ is linearly dependent
on $q$ and has been estimated in Ref.~\onlinecite{marsman}.  With the
slightly smaller simulation cell used in this work, the $\phi_q$ corrections
are predicted to be 3.85 and 3.81~eV for 1.00 and 0.99~g/cc water
density.\cite{note2}
Unlike classical force field calculations, the isolated ion
Li$^{q +}$ carries a non-zero energy.  Thus we subtract
$ (d H_{\rm bare~ion}(q) /dq )_q$ from Eq.~\ref{eq1}. 

Unless otherwise noted, the Li$^+$ thermodynamic integration protocol
(e.g., the sampling interval, subtraction of bare ion energies) is
applied to all other ions.

\subsection{Cl$^-$ thermodynamic integration}
$\Delta G_{\rm hyd}$ for Cl$^-$ requires a different TI procedure.  Unlike
the Li$^+$ PP without explicit $1s$ electrons, scaling the {\tt VASP} Cl$^-$ PP to
zero also involves removing 8 electrons.  While it is possible to
alchemically perturb Cl$^-$ to Ar, this TI route is not directly applicable
for multi-atom anions.  Instead, we first use TI to ``grow'' a 
non-polarizable {\it classical} force field (FF) \cite{garde} Cl$^-$
with a negative point charge and a Lennard-Jones interaction\cite{garde} with
the oxygen sites of PBE water.  This can be regarded as a QM/MM simulation,
but with the solvent (not solute) treated quantum mechanically.
Then we use a one-step free energy perturbation (FEP) procedure, 
\begin{eqnarray}
&& \beta [\Delta G ({\rm PBE}) -\Delta G({\rm FF})]  \nonumber \\
	&=& -\log \langle \exp[ -\beta (H({\rm PBE})
		- H({\rm FF})] \rangle_{\rm FF} , \label{fep}
\end{eqnarray}
to estimate the PBE Cl$^-$ $\Delta G_{\rm hyd}$.  As long as the hydration
structures of the classical and PBE ion in PBE water are similar, this method
can be generally and accurately applied to multi-atom anions or cations, as well
as PP's like the {\tt VASP} PAW PBE Na$^+$ whose interaction with water exhibits
anomalies when the PP is scaled continuously to zero (Fig.~\ref{fig1}).
If there are partial positive point charges in the classical force field,
however, the DFT valence electrons may collapse onto those atomic sites,
and pseudopotentials that repel electrons may be needed to prevent such
a collapse.

\subsection{Ag$^+$ and Ni$^{2+}$ thermodynamic integration}

The {\tt VASP} PBE pseudopotentials used for Ag and Ni contain 11 and 16 electrons,
respectively.  When the number of electrons in 32-water simulation
cell is fixed at $(32\times 8+11-q)$ and $(32\times 8+16-q)$ in AIMD
trajectories, our maximally localized Wannier function analyses\cite{wannier}
reveal that $(11-q)$ and $(16-q)$ electrons remain localized on Ag and Ni,
respectively.  This indicates that Ag$^{q+}$ and Ni$^{q+}$ species exhibit no
tendency to eject excess electrons into water,\cite{note} and
the partially charged ions are preserved within a $\lambda$-path that
vary the total number of electrons in the system.  Hence we simply use
the number of electrons as the order parameter, $\lambda$, analogous
to Refs.~\onlinecite{anatole1},~\onlinecite{zeng}, and~\onlinecite{seealso}.
$\frac{d H(q)}{dq}$ is simply computed by adding and subtracting
0.025~electrons to the simulation cell and performing a finite difference.
The exceptions are
Ag$^+$ (where we compute the difference between Ag$^+$ and Ag$^{0.95+}$);
Ni$^+$ (Ni$^+$ and Ni$^{1.05+}$); and Ni$^{2+}$ (Ni$^{1.95+}$ and Ni$^{2+}$).
As we subtract the bare ion contribution at each $q$, the expression
($\langle \frac{d H(q)}{dq}\rangle  -  \frac{d H_{\rm bare~ion}(q) }{dq} )$
should reflect purely solvent-induced effects.

For Ag, spin-polarized PBE calculations are adequate.  In contrast,
spin-polarized PBE-based AIMD simulations of Ni$^{q+}$ in water underestimate
the gap between the highest occupied (HOMO) and lowest unoccupied (LUMO)
molecular orbitals.  This occurs because PBE severely underestimates
exchange interactions in the localized $3d$ orbitals, leading to
near degeneracies in intermediate-$q$ Ni$^{q+}$ $d$-shell orbitals and
slow numerical convergence of the electronic structure at each Born-Oppenheimer
AIMD time step.  We have therefore applied the DFT+U technique\cite{dftu} to
the Ni $3d$ orbitals to generate AIMD trajectories with which we
evaluate Eq.~\ref{eq1} using only the PBE functional.  Originally devised for
solid state applications, DFT+U has recently been adapted for molecular
systems and even used in AIMD settings.\cite{leung1,dftu_aimd1}
$U$ is set at 4.0~eV to yield a 15.7~eV gas phase Ni$^{2+}$ binding energy in a 
Ni$^{2+}$(H$_2$O)$_6$ cluster.  This is the value predicted
using the B3LYP hybrid functional\cite{b3lyp} and a 6-311+G(d,p)
basis.\cite{basis}  
Using DFT+U generated geometries for PBE $\Delta G_{\rm hyd}$ is justified
because, in the gas phase, the PBE functional and DFT+U predict
optimized Ni$^{2+}$(H$_2$O)$_6$ geometries which are nearly identical.

\section{Results}
 
\subsection{Li$^+$ hydration free energy}
 
\begin{figure}
\centerline{\hbox{\epsfxsize=3.00in \epsfbox{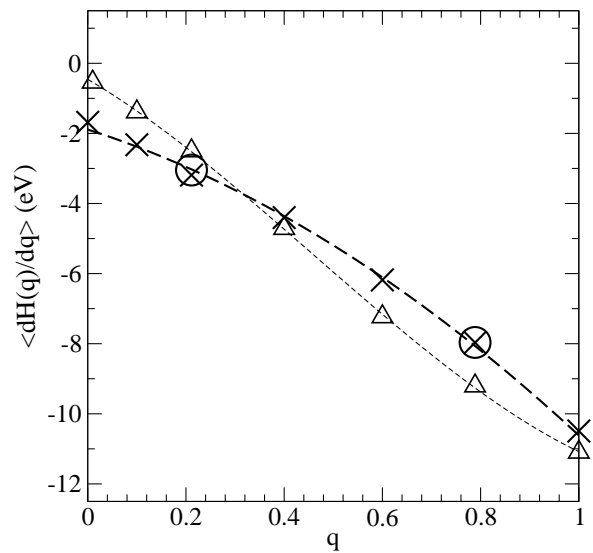}}}
\caption[]
{\label{fig2} \noindent
$\langle dH(q)/dq \rangle_q$ for Li$^{q+}$ as $q$ varies.  
The bare ion contributions, Ewald corrections, and
electrostatic potential shift due to the quadrupole moment
have been subtracted.  Crosses: 32 H$_2$O, 1.00 g/cc; circles: 64
H$_2$O, 1.00 g/cc; triangles, same as crosses but are for SPC/E water.
The dashed lines are cubic least-squared fits to the crosses and triangles.
}
\end{figure}

Figure~\ref{fig2} plots $\langle d H(q)/dq \rangle_q$ as $q$ varies
after subtracting contributions from Ewald images,\cite{hummer_mono}
the quadrupole or spherical second moment contribution $q\phi_q$,\cite{marsman}
and the energies of the
bare Li$^{q+}$.  $\langle d H(q)/dq \rangle_q$ computed using
32- and 64-H$_2$O simulation cells at 1.00 g/cc H$_2$O density
are in good agreement at $q=0.21$ and $q=0.79$.  Using a 2-point Gaussian
quadrature, $\Delta G_{\rm hyd}$ for the two cells integrate to -128.6 and
-126.7~kcal/mol, respectively (Table~\ref{table1}).  Splitting the data
into four segments, the standard deviations in these $\Delta G_{\rm hyd}$
are found to be 1.1 and 0.5~kcal/mol, respectively.\cite{note5}
Thus the two cell sizes exhibit $\Delta G_{\rm hyd}$ approximately within
numerical uncertainties of each other, showing that the finite system
size effect is small for AIMD after applying the Ewald correction,
as is the case with classical force field simulations.\cite{hummer_mono,na8} 
A dielectric continuum estimate would suggest that, after adding the
leading order ($1/L$) Ewald correction, the 32-water simulation cell
result is already converged to the infinite dilution limit to within
1~kcal/mol (Ref. 9).

\begin{figure}
 \centerline{\hbox{\epsfxsize=2.50in \epsfbox{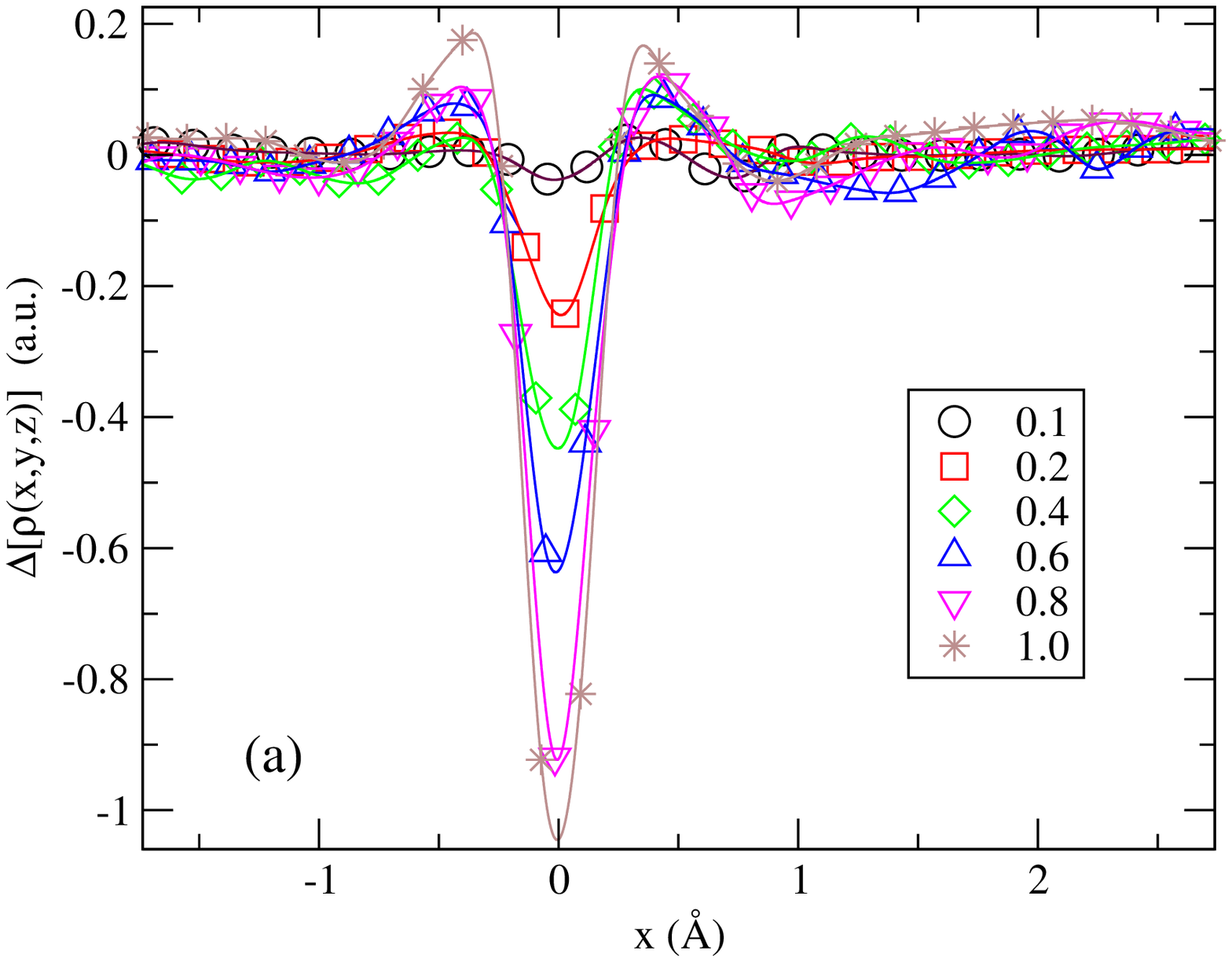}}}
 \centerline{\hbox{\epsfxsize=1.75in \epsfbox{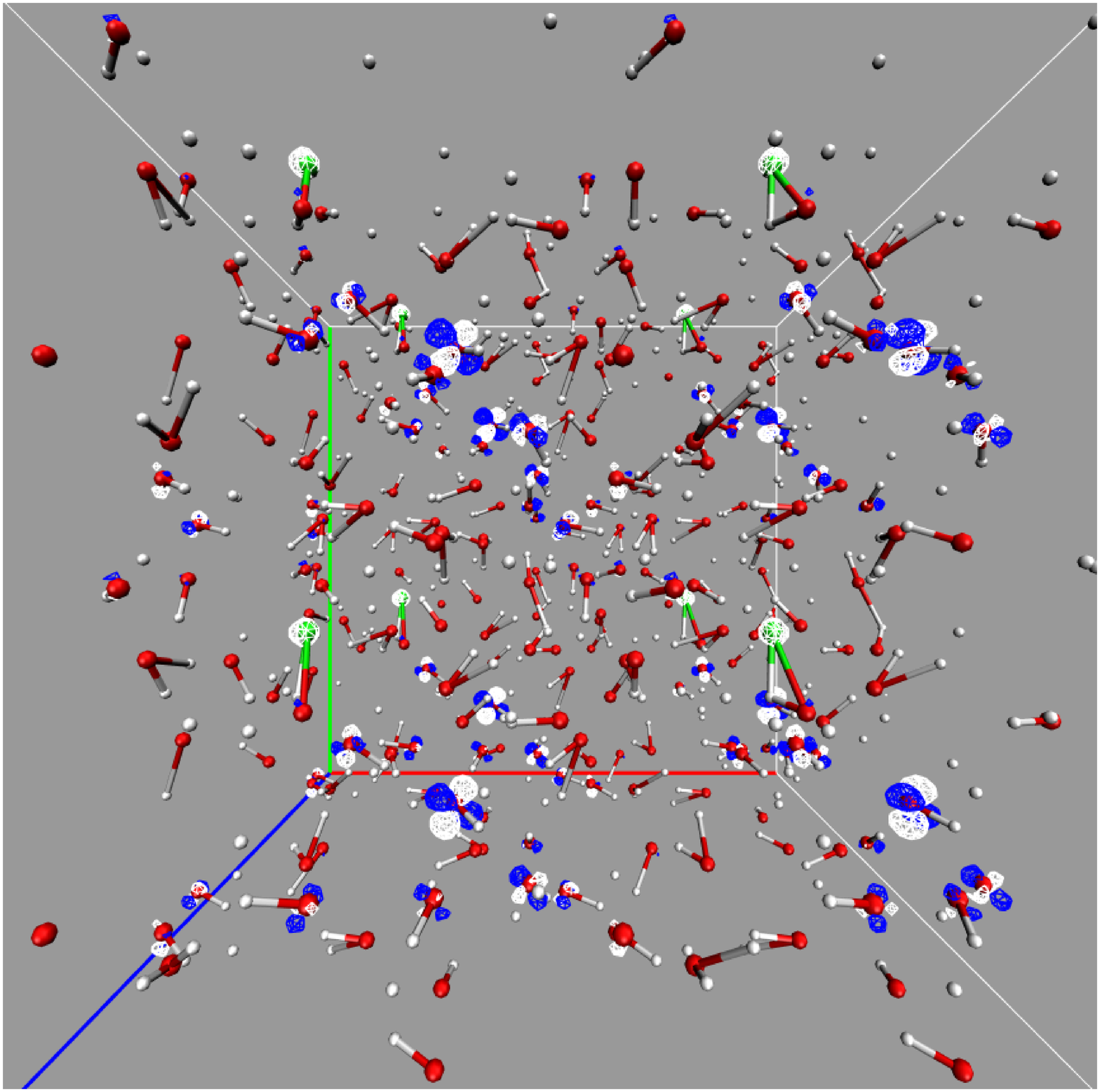} 
                   \epsfxsize=1.75in \epsfbox{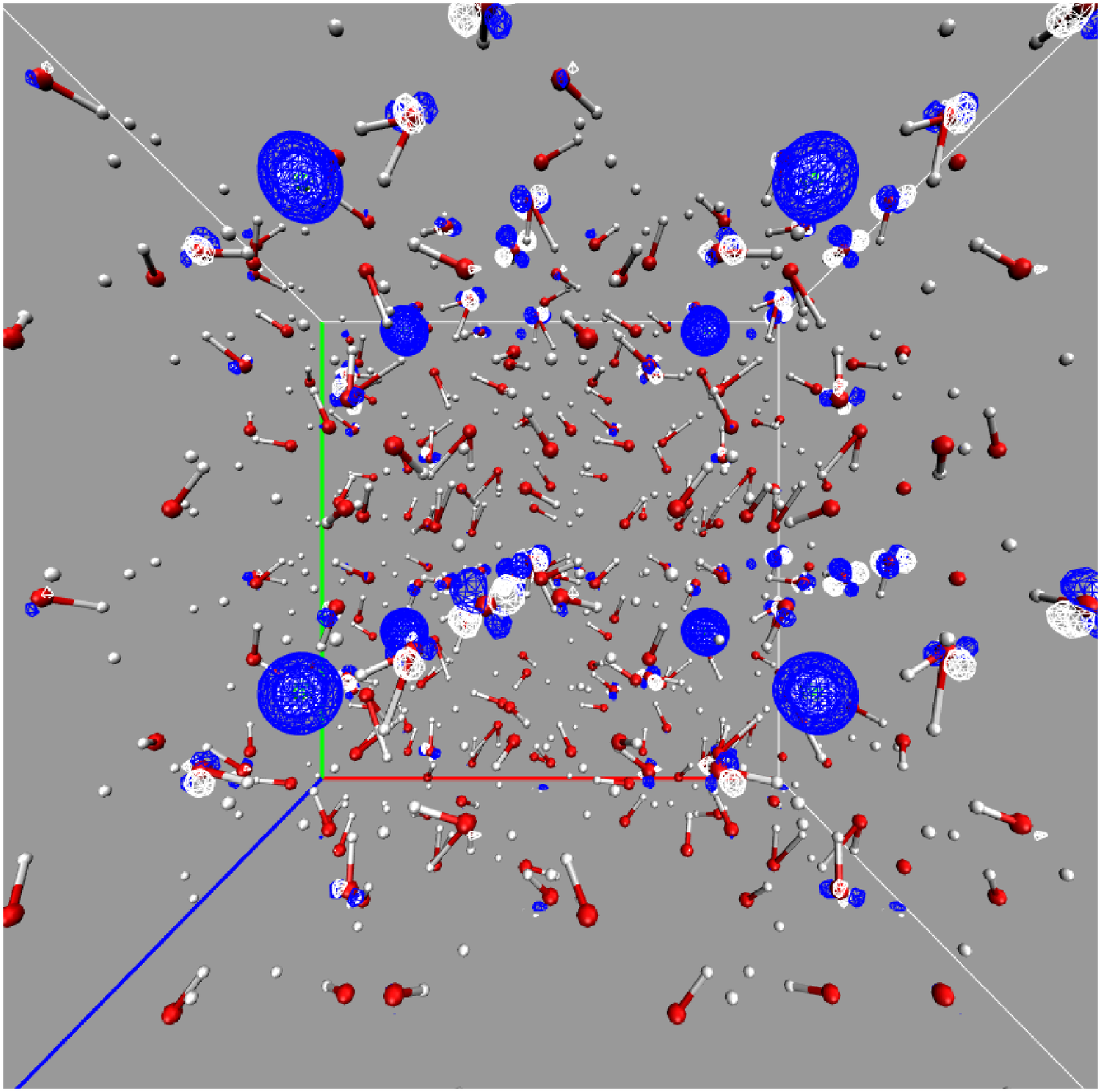}}}
\centerline{\hbox{ (b) \hspace*{2in} (c)}}
  \centerline{\hbox{\epsfxsize=1.75in \epsfbox{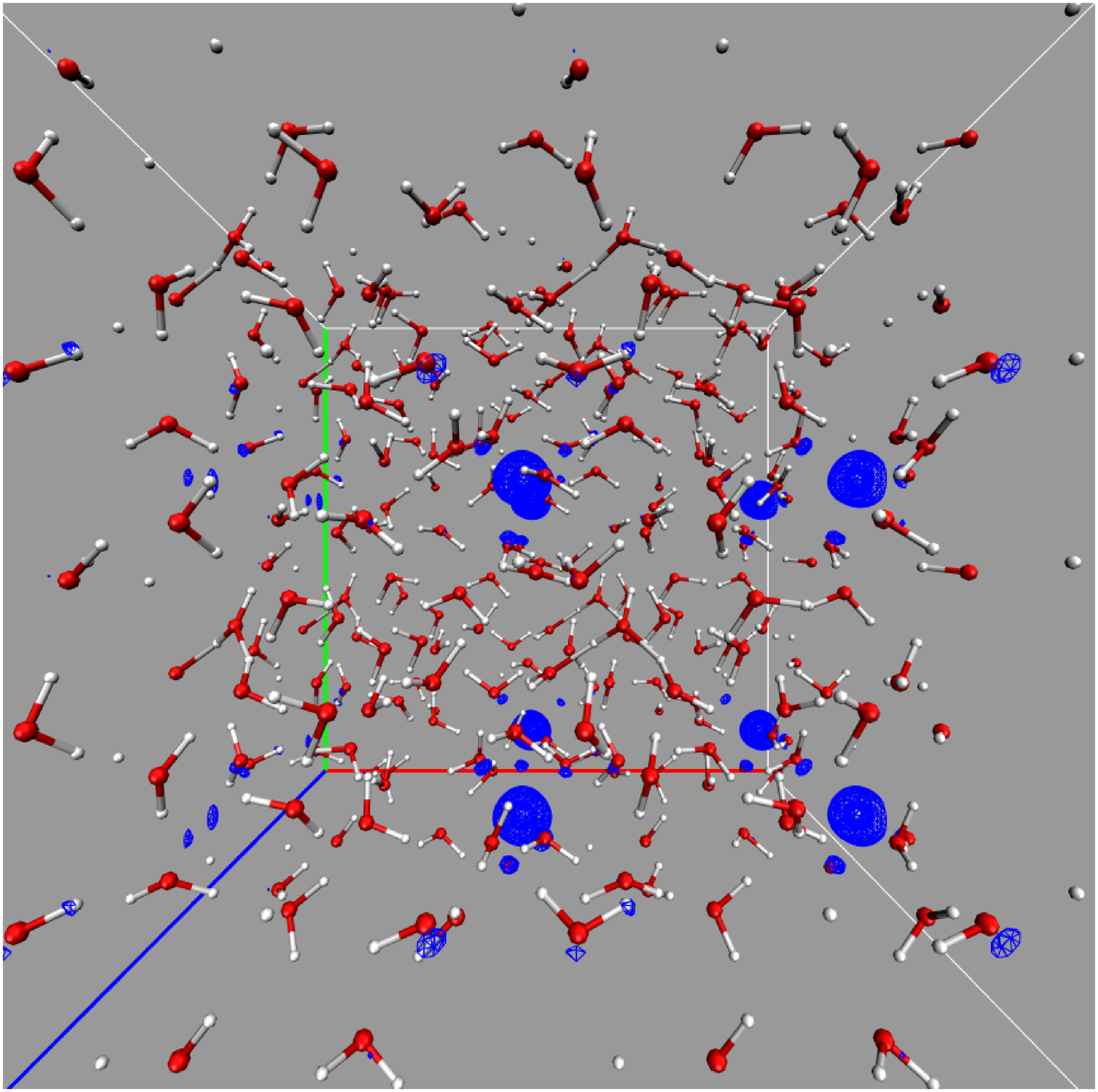}}}
 \centerline{\hbox{ (d)}}
\caption[]
{\label{fig3} \noindent
(a):
Integrated changes in electron density,
$\Delta (x) = \int dy dz [\rho(x,y,z)_{n} - \rho(x,y,z)_c]$,
as a function of spatial coordinate $x$ for the various values of $q$.
$\rho_n$ and $\rho_c$ are the densities for the neutral and the charged
systems, respectively.  All charged species, Li$^{q+}$
have been shifted to $x$ = 0.
Symbols correspond to actual grid-points, the continuous
lines are cubic interpolations.
(b)-(d):
Isosurface plots of the electron density difference,
$\rho(x,y,z)_{n} - \rho(x,y,z)_c$ (iso-value = $\pm$ 0.01 a.u.,
white $\le$ 0, blue $\ge$ 0), for $q$ =0.1, 0.6, and 1.0.
Periodic boundary conditions apply; the prominent,
8 blue spheres represent the (periodically replicated)
changes in Li$^{q+}$ densities, and some changes
in water dipole moments are apparent too.
See Sec.~\ref{anatole} for technical details.
}
\end{figure}

For illustrative purposes, we also display in Figure~\ref{fig3} 
the Li-ion growth-induced changes of the total electron density 
integrated over the $x$- and $y$-coordinates.  From inspection of this
change arising from the presence of the increasingly
charged ion one can conclude that, as expected, the attraction of
electrons toward the ion increases as the charge approaches +1.0.
The isosurface plots support a similar conclusion.
For small values of $q$, changes in density occur throughout
the system.  As $q$ approaches its final value, however,
the drastic increase in electronic density at the ion position due to
increasingly polarized water (Fig.~\ref{fig3}a) is hidden behind the
large sphere of depleted density.  This large sphere comes about
because we have subtracted the electron density of a neutral
helium atom from that of the Li$^+$ pseudopotential.

\begin{figure}
\centerline{\hbox{\epsfxsize=3.00in \epsfbox{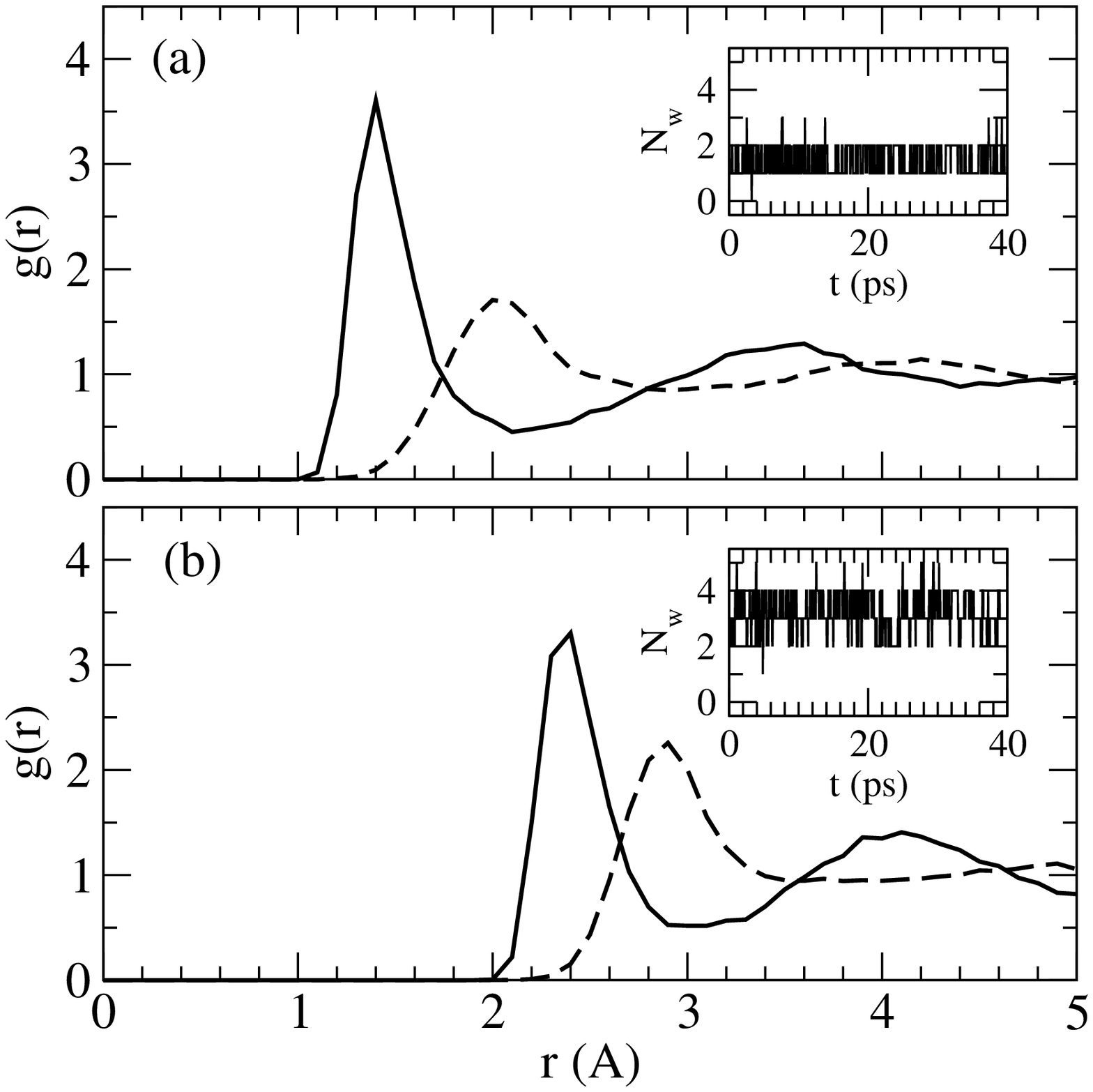}}}
\caption[]
{\label{fig4} \noindent
Pair correlation functions $g(r)$ between Li$^{q+}$ and
the O (solid line) and H (dashed line) sites of H$_2$O molecules.
(a) $q$=0.21; (b) $q$=0.79.  The
instantaneous hydration numbers are depicted
in the insets.
}
\end{figure}

Figure~\ref{fig4} depicts the pair correlation functions $g(r)$ 
between Li$^{q+}$ and the O and H sites in H$_2$O.  Recall that the entire
{\tt VASP} PBE Li$^+$ pseudopotential, including the long-range coulomb
and the short-range Pauli-exclusion contributions, is scaled with $q$.
Hence, at small $q$, the most probable Li$^{q+}$-O$_{\rm water}$ distance
is much reduced from the $q=1$ case.  Nevertheless, we have verified
that negligible electron density resides near the Li$^{q+}$ nuclei,
indicating that Li$^{q+}$ does behave like a partially charged ion in water.
The insets depict the instantaneous hydration numbers $N_w$,
computed at each time step by integrating each $g_{\rm Li-O}(r)$ to its
first minimum.  For $q=0.21$, $N_w$ averages only to 1.5 and experiences rapid
temporal fluctuations.  Despite this, $g_{\rm Li-O}(r)$ still exhibits a
high peak value because the scaled Li$^{q+}$ has such a small radius.
At $q=0.79$, $N_w = 3.5$, approaching the $N_w=4 $ AIMD value reported for
Li$^+$.\cite{li_rempe}
 
\begin{figure}
\centerline{\hbox{\epsfxsize=3.00in \epsfbox{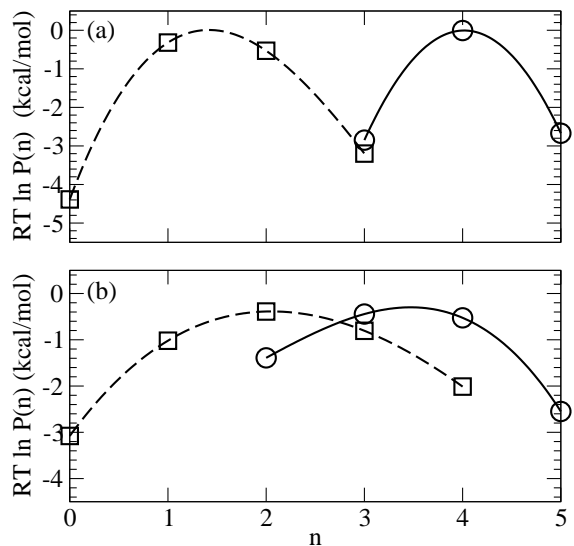}}}
\caption[]
{\label{fig5} \noindent
Logarithm of the probability ($P_n$) of instantaneous hydration numbers
($n$) multiplied by thermal energy, in units of kcal/mol.
(a) Li$^{q+}$; (b) Ag$^{q+}$.  Squares and
dashed lines: $q=0.2$; circles and solid lines: $q=1.0$.
$n$ is determined by counting all water oxygen atoms within 2.08,
2.75, 2.90, and 2.92~\AA\, of the four ions, respectively.
These distances are determined by locating ithe first minimum in the
ion-water $g(r)$.
}
\end{figure}

Figure~\ref{fig5} depicts the logarithm of the distributions of
instantaneous hydration numbers for Li$^{0.2+}$ and Li$^+$.  
In conjunction with low order $m$ $\langle d^m H(q)/dq^m\rangle$ derivatives,
hydration number distributions at the TI end-points
can in princple be used to predict the hydration free energy
using a single AIMD trajectory at $q=0$ or $q=1$.\cite{garcia}  Since
we have avoided $q=0$ and the finite differences applied in our
implementation may not be accurate for $m>1$, we have not attempted
to estimate $\Delta G_{\rm hyd}$ with high order derivatives,
but have used 2 or 6 $q$ values to evaluate $\Delta G_{\rm hyd}$.
Note that, using the quasi-chemical theoretical framework, hydration number
distributions of a solute can be used directly to estimate
hydration free energies,\cite{quasichem}, as demonstrated in recent
works.\cite{sabo,quasichem,paliwal}
Furthermore, such distributions are of intrinsic interest
and can lend useful comparison with those predicted using classical
force field simulations.  See also Ref.~\onlinecite{hummer_mol_sim}
for other methods devised to reduce the number of $q$-value integrands
needed to perform TI calculations.

We next investigate the accuracy of the 2-point TI quadrature by further
sampling $\langle dH(q)/dq\rangle_q$ at $q$=0.1, 0.4, 0.6, 1.0 in addition
to 0.21 and 0.79 in a simulation cell.
This denser grid allows an approximate 6-point trapezoidal rule
integration after we extrapolate $\langle dH(q)/dq\rangle_q$ to $q$=0.0.
Figure~\ref{fig2} shows that $\langle dH(q)/dq\rangle_q$ is almost
linear for a large, intermediate $q$ range except near $q$=0 and $q$=1.
This is in qualitative agreement with SPC/E model
predictions\cite{classical,na8} which we also compute 
for a 32-water simulation cell and depict in Fig.~\ref{fig2}.
The deviation from linearity at $q=0$ is well-reproduced with a cubic fit
for both AIMD and SPC/E $\langle dH(q)/dq\rangle_q$.  Table~\ref{table1}
confirms that the 2-point and 6-point formulas yield $\Delta G_{\rm hyd}$
within 0.3~kcal/mol of each other --- well within the numerical uncertainties
of the simulations.  Henceforth we will report the 6-point value of
$\Delta G_{\rm hyd}$ $=-128.3 \pm 0.9$~kcal/mol for Li$^+$.

This success of the 2-point formula appears however somewhat fortuitous.
One would not 
{\it a priori} expect this quadrature to be accurate for Li$^+$ because of
the large changes in effective Li$^{q+}$ radius (Fig.~\ref{fig4}).
The classic Born hydration free energy
formula, based on a dielectric continuum description of the solvent,
predicts $\Delta G_{\rm Born} \propto q^2/(2a) (1-1/\epsilon)$ at a fixed
ionic radius $a$.  It is quadratically dependent on $q$ when $a$ is held
constant.  In non-polarizable classical force field $\Delta G_{\rm hyd}$
simulations, the Lennard-Jones radius of the ion is also held fixed while
the ionic charge varies.  The constant radius thus seems crucial to the
accuracy of the 2-point Gaussian quadrature, which is exact only if
$\langle dH(q)/dq\rangle_q$ is linear in $q$.  Despite this, the 2-point
formula will be shown to be accurate for the AIMD $\Delta G_{\rm hyd}$
associated with Li$^+$, Ag$^+$, and Ni$^+$ $\rightarrow$
Ni$^{2+}$ considered in this work.  It appears less accurate for Cl$^-$,
unlike SPC/E-based Cl$^-$ $\Delta G_{\rm hyd}$ calculations.
The fact that the radius of Li$^{q+}$ (and to some extent, other ions)
changes with $q$ in our DFT calculations also explains the discrepancy
between AIMD and SPC/E $\langle dH(q)/dq \rangle_{q=0}$ values.

To compare AIMD predictions with experimental data,
$\Delta G_{\rm Marcus}+q\phi_q^{\rm SPC/E}-2C^{(0)}$~kcal/mol is found to be
-137.0~kcal/mol,\cite{marcus} while
$\Delta G_{\rm Tiss}-q\phi_d^{\rm SPC/E}-C^{(0)}$~kcal/mol
=-133.2~kcal/mol\cite{tiss} (Table~\ref{table1}).  These values are similar
to the SPC/E $\Delta G_{\rm hyd}$ for Li$^+$, and are 8.7 and 4.9~kcal/mol
higher than the 6-point AIMD prediction for a 32 H$_2$O simulation cell,
respectively.  The discrepancies with AIMD predictions may be due to numerical
noise, PBE functional inaccuracies, or systematic uncertainties arising from
the treatment of $|e|\phi$.  Indeed, the discrepancy between SPC/E-augmented
experimental values listed by Marcus\cite{marcus} and Tissandier
{\it et al.}\cite{tiss} can also be taken as a measure of surface
potential-related systematic ambiguity.  This issue will be interrogated
in the next subsection when we consider the anion Cl$^-$.

An optimal study of hydration free energy would include also the
changes in water density due to the presence of salt cations and
anions or water confinement inside nanopores.  We have therefore
examined the effects of reducing the water density to 0.97~g/cc.
This small reduction in water density corresponds to the activity
of water at 0.1~M ion concentration, which is the typical concentration
of K$^+$ ions in the cytoplasm of skeletal muscle cells and the typical
concentration of Na$^+$ and Cl$^-$ ions outside cells.\cite{hille}.
Table~\ref{table1} shows that the small effect on $\Delta G_{\rm hyd}$ 
due to water density changes is within the numerical uncertainty.
This weak dependence is consistent with
quasi-chemical theory analysis\cite{varma,quasichem} where contributions to
$\Delta G_{\rm hyd}$ are separated into inner hydration shell and outer
shell contributions.  In the ``cluster'' implementation of the
theory,\cite{sabo} the former can be determined from gas phase cluster
calculations scaled by water density, while the latter depends on the water
dielectric constant, which is relatively independent of H$_2$O density.
As pointed out by Varma and Rempe,\cite{varma} since the dependence of
free energies on water concentration is logarithmic, large 
changes in water density are required before there is an effect
on $\Delta G_{\rm hyd}$.

\subsection{Cl$^-$ hydration free energy}

\begin{figure}
\centerline{\hbox{\epsfxsize=3.00in \epsfbox{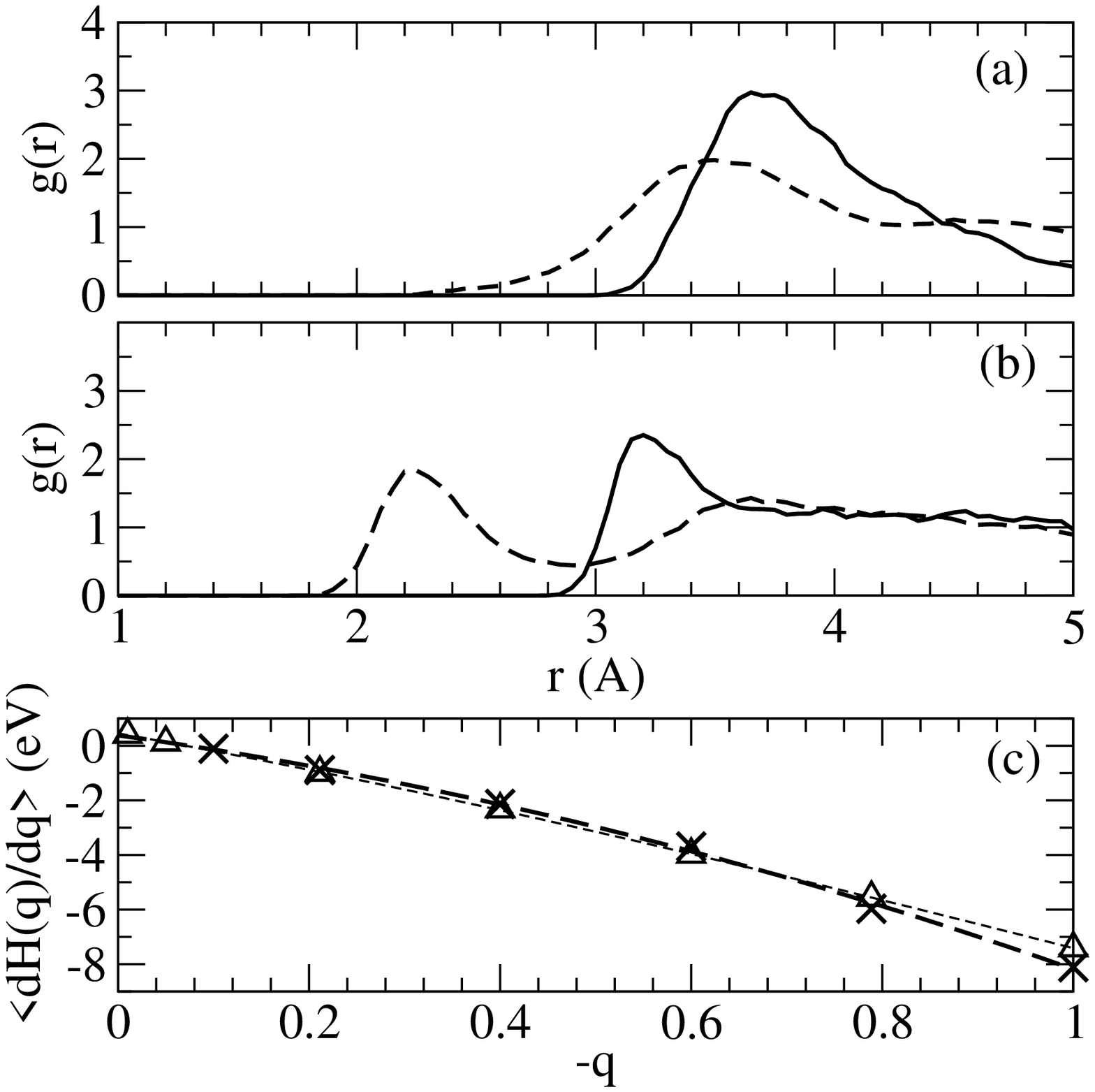}}}
\caption[]
{\label{fig6} \noindent
(a),(b) Pair correlation functions $g(r)$ between classical force field
Cl$^{q-}$ and the O (solid line) and H (dashed line) sites of PBE H$_2$O
molecules.  (a) $q$=0.21; (b) $q$=0.79.  
(c) $\langle dH(q)/dq \rangle_q$ for classical force field Cl$^{q-}$
as $q$ varies.   Crosses and triangles are for AIMD and classical force
field treatments of water in 32-H$_2$O simulation cells.
The bare ion contributions, Ewald corrections, and electrostatic potential
shift due to the quadrupole moment have been subtracted. 
The dashed lines are cubic least-squared fits.
}
\end{figure}

Figures~\ref{fig6}a~and~b depict the $g(r)$ between 
the classical force field Cl$^{q-}$ (henceforth FF-Cl$^{q-}$)
and the oxygen and proton sites of H$_2$O molecules at two $q$
values.  At $q$=0.21 (or even $q$=0.4), FF-Cl$^{q-}$ is predominantly a
hydrophobic sphere that excludes both O and H from its vicinity.
Due to the sheer size of the Lennard-Jones sphere that represents
Cl$^{q-}$, this solute is seen to substantially disrupt
the water structure around it in the 32-H$_2$O simulation cell.
Thus, in panel (b), the Cl-O $g(r)$ has dropped below 0.5 density units
at $r\sim 5$\AA\, --- unlike the case for Li$^{q+}$ at small
$q$ (Fig.~\ref{fig4}a).
At $q$=0.79, the ion forms hydrogen bonds with water;
its $g_{\rm Cl-H}(r)$ exhibits a peak at $r=2.2$\AA.
At $q$=1 (not shown), we obtain
a FF-Cl$^{q-}$ hydration number of $N_w$=5.4, in good
agreement with full AIMD simulations of PBE Cl$^-$ in PBE
water.\cite{cl_aimd1,rigid}

Figure~\ref{fig6}c depicts the variation of $\langle dH(q)/dq \rangle_q$
FF-Cl$^{q-}$ in PBE water as $q$ varies.\cite{note4} To obtain
$\Delta G_{\rm hyd}$ for the PBE Cl$^-$ ion, we further apply Eq.~\ref{fep}
to configurations sampled 0.1~ps apart along the AIMD trajectory.
The differences between the instantaneous
potential energies for FF-Cl$^{-}$
and PBE Cl$^{-}$ are found to be almost constant with an estimated standard
deviation of 0.15~kcal/mol.  This indicates that FF-Cl$^-$ is an excellent
reference for the PBE Cl$^-$.  After a cubic polynomial extrapolation to $q$=0
and applying a 6-point integration formula, $\Delta G_{\rm hyd}$ for the PBE
Cl$^-$ integrates to -76.6$\pm$0.4~kcal/mol (Table~\ref{table1}).  A 2-point
Gaussian quadrature formula yields -79.0$\pm$0.8~kcal/mol.  As the latter is
only exact for linear $\langle dH(q)/dq \rangle_q$, deviation from linearity in
Fig.~\ref{fig6} indicates that a denser grid may be needed despite the constant
radius of the FF-Cl sphere.  This slight non-linearity 
is apparently due to water polarizability; corresponding 6-point
and 2-point SPC/E calculations in 32-water simulation cells yield
indistinguishable results.  As
$\langle dH(q)/dq \rangle_q$ is well-fitted to a cubic polynomial in
$q$ and the trapezoidal integration rule is accurate for cubic
polynomials, however, Fig.~\ref{fig6}c strongly suggests that an
integration formula higher order than the trapezoidal rule is not needed.
Henceforth we report the 6-point TI value.

Two post-processing corrections for $\Delta G_{\rm hyd}$, unnecessary
for Li$^+$, need to be included here.  (1) While the Li$^+$ PP is globally
shrunk to zero, at $q$=0 FF-Cl$^{q-}$ remains a Lennard-Jones sphere that
displaces water.  This gives rise to an entropic or ``packing''
penalty; the contribution is estimated to be 4.0~kcal/mol using SPC/E
water model simulations.  (2) Simulation cell size effects are more
significant for Cl$^-$ than for Li$^+$, presumably because of the size of
the Cl$^{q-}$ sphere at small $q$ (Fig.~\ref{fig6}a).  When we perform
purely classical force field simulations of
a Cl$^-$ ion in SPC/E water, we find that a 32-H$_2$O simulation 
cell overestimates $\Delta G_{\rm hyd}$ by 3.3~kcal/mol compared
to a 255-H$_2$O cell.  This discrepancy is much larger than the numerical
uncertainty.  In contrast, these two cell sizes yield Li$^+$
$\Delta G_{\rm hyd}$ that are within about 1~kcal/mol.
The simulation cell size dependence has been estimated
using a dielectric continuum approach in Ref.~\onlinecite{na8}.
Assuming AIMD exhibits Cl$^-$ packing penalty and simulation cell size
dependence similar to classical force field MD, we add a 7.3~kcal/mol correction
to the AIMD result.  The corrected AIMD Cl$^-$ $\Delta G_{\rm hyd}$ is
listed in Table~\ref{table3}.  It is within 0.4~kcal/mol of
$\Delta G_{\rm Tiss} - q\phi_d^{\rm SPC/E}-C^{(0)}$~kcal/mol, and
overestimates the magnitude of $\Delta G_{\rm Marcus}
+ q\phi_q^{\rm SPC/E}-2C^{(0)}$~kcal/mol by 4.0~kcal/mol.


Adding $\Delta G_{\rm hyd}$ of oppositely charged monovalent ions
eliminates the systematic uncertainty due to surface potential contributions.  
The combined $\Delta G_{\rm hyd}$ for Li$^+$ and Cl$^-$ are within 4.7 and
5.3~kcal/mol of experimental data quoted in Table~\ref{table3}
respectively.\cite{marcus,tiss} they underestimate those values only
by about 2.3 and 2.6\%.  This sum, derived from Marcus\cite{marcus}
and Tissandier {\it et al.},\cite{tiss} are within 0.6~kcal/mol of each
other, unlike in the cases of the isolated Li$^+$ and Cl$^-$ ions
where the two adjusted experimental data sets disagree by 3.8 and
4.4~kcal/mol, respectively.  This suggests that the rather large, 8.7~kcal/mol
discrepancy between AIMD $\Delta G_{\rm hyd}$ and Marcus' data for Li$^+$
is partly due to the assignment of the SPC/E $\phi_q$ contribution
to the surface potential.   In contrast, Tissandier {\it et al.}'s
data for the isolated ions are in substantially better agreement
with AIMD $\Delta G_{\rm hyd}$ for both ions, suggesting that
augmenting $\Delta G_{\rm Tiss}$ with SPC/E $\phi_d$ is a reasonable
approximation.

\subsection{Ag $\rightarrow $ Ag$^+$}
 
\begin{figure}
\centerline{\hbox{\epsfxsize=3.00in \epsfbox{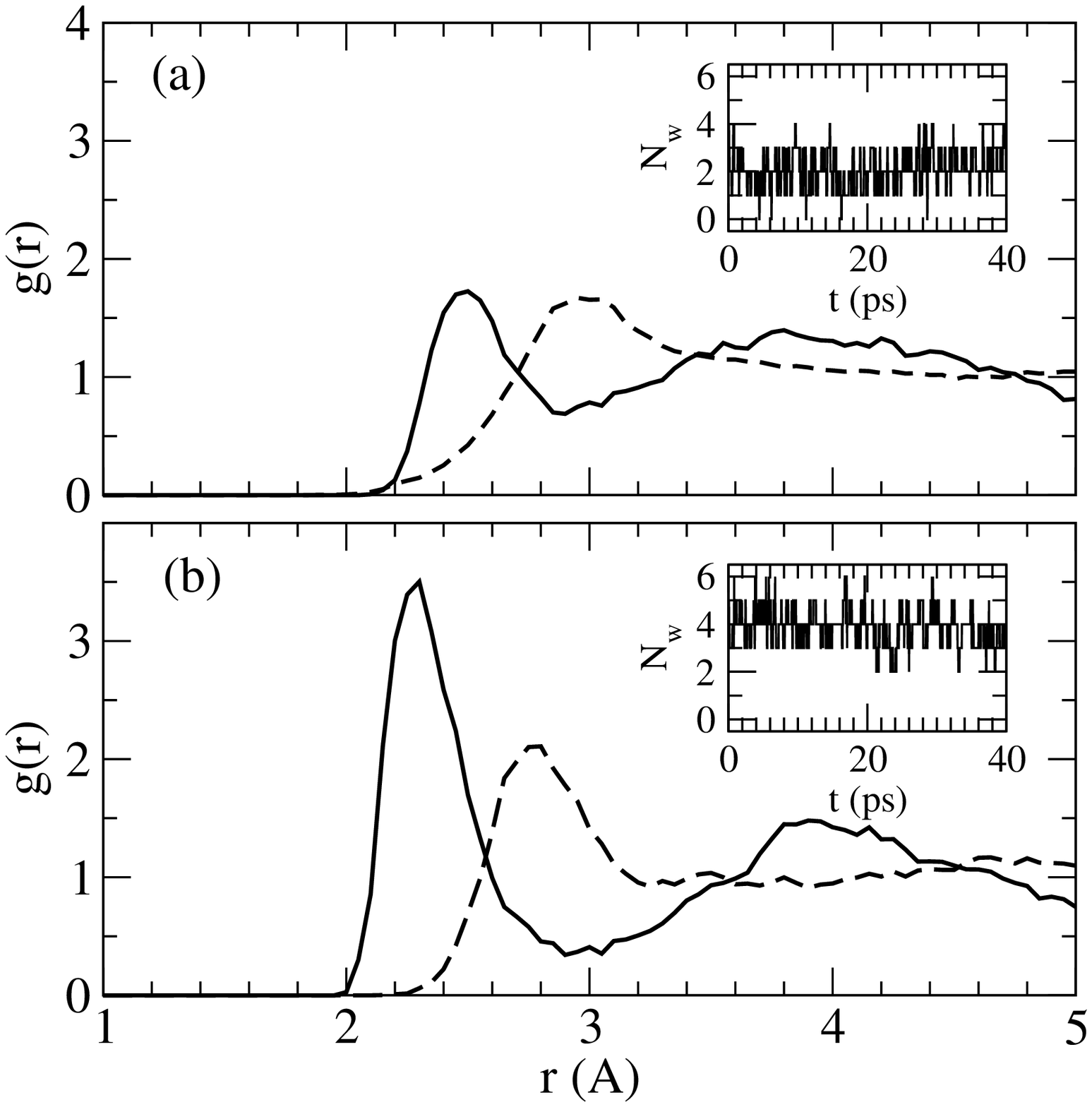}}}
\caption[]
{\label{fig7} \noindent
Pair correlation functions $g(r)$ between Ag$^{q +}$ and
the O (solid line) and H (dashed line) sites of H$_2$O molecules.
(a) $q$=0.21; (b) $q$=1.00.  The
instantaneous hydration numbers are depicted
in the insets.
}
\end{figure}

In Fig.~\ref{fig7}, Ag-O$_{\rm water}$ and Ag-H$_{\rm water}$ $g(r)$
are depicted for two selected values of $q$.  Unlike Li, the Ag atomic
core is not scaled with $q$, and Pauli repulsion
ensures that no water molecule penetrates the Ag core region.
Thus the $g(r)$ is not sharply structured at small $q$, and Ag$^{q+}$
resembles a hydrophobic sphere as $q$ decreases.  For both $q$ points,
H$_2$O in the first hydration shells are highly labile; see the insets. 
The Ag$^{+}$-H$_2$O $g(r)$ (Fig.~\ref{fig7}b) yields a first shell hydration
number $N_w$=3.4.  The instantaneous hydration number distribution is
depicted in Fig.~\ref{fig5}.  This $N_w$ is qualitatively similar to
the $N_w$=4.0 computed using AIMD and another exchange correlation
functional.\cite{sprik_ag} Both these AIMD $N_w$ values are in good
agreement with experiments.\cite{ag_expt1,ag_expt2}  
In contrast, a recent classical
force field model with parameters fitted to quantum chemistry calculations
has reported $N_w=6$.\cite{ag_ff} With the corrections (A)-(B) discussed
earlier, a 6-point trapezoidal rule integration, and a 1.6~kcal/mol
packing correction estimated using classical force field simulations,
we obtain $\Delta G_{\rm hyd}$=-119.8$\pm 0.4$ kcal/mol.
This magnitude is 6.4~kcal/mol smaller than $\Delta G_{\rm Marcus}
+ q \phi_q^{\rm SPC/E}-2C^{(0)}$~kcal/mol (Table~\ref{table4}).\cite{marcus}
The sum of AIMD Ag$^+$ and Cl$^-$ $\Delta G_{\rm hyd}$, however, underestimates
the experimental data\cite{marcus} by only 2.4~kcal/mol, or by 1.2~\%.

\subsection{Ag$^+$ + Ni$^+$ $\rightarrow$ Ag + Ni$^{2+}$}

\begin{figure}
\centerline{\hbox{\epsfxsize=4.00in \epsfbox{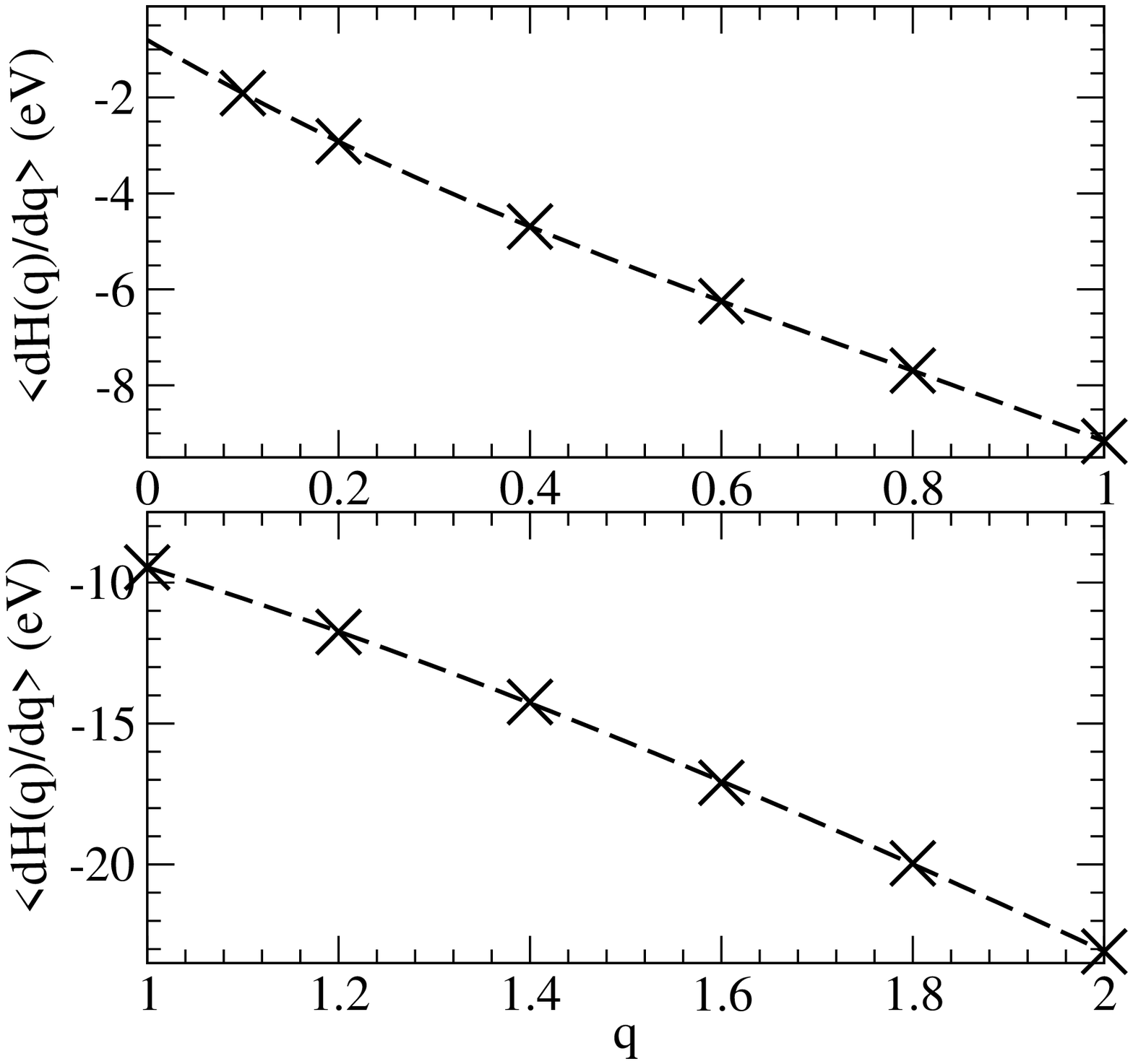}}}
\caption[]
{\label{fig8} \noindent
$\langle d H(q)/dq \rangle_q$ for Ag$^{q+}$ and Ni$^{q+}$ as $q$
varies.  The bare ion contributions, Ewald corrections, and global
shift in the electrostatic potential due to the quadrupole
moment have been accounted for.  The dashed lines are cubic least-squared fits.
}
\end{figure}

The details of Ni$^{q+}$ hydration will be described elsewhere.\cite{future}
Here we focus on the change in $\Delta G_{\rm hyd}$ as Ni$^+$ loses an electron.
We use the PBE functional to compute $\langle d H(q)/dq \rangle_q$ at 0.1~ps
intervals along the DFT+U AIMD trajectory with $U$=4~eV.  Figure~\ref{fig8}b
shows that $\langle d H(q)/d q \rangle_q$ is fairly linear as $q$ varies.  
With a 6-point trapezoidal
rule integration, Eq.~\ref{eq1} yields a change in $\Delta G_{\rm hyd}$ of
-365.5$\pm 1.0$~kcal/mol.  A 2-point integration predicts a similar
-363.4$\pm 2.4$~kcal/mol.  Unlike the calculations for Li$^+$ and Ag$^+$, this
system benefits from the fact that at ``$\lambda$''=$(q-1)$=$0$, Ni$^+$ is
still highly charged, and larger statistical uncertainty at small $q$ is
avoided.
Nevertheless, due to the slower water dynamics around the more highly
charged Ni$^{q+}$ ion, sampling correlation times may be longer and
our error bars for Ni$^{2+}$ may be underestimated.

The electrochemical half cell reaction free energy consists of the change
in $\Delta G_{\rm hyd}$ plus the ionization potential (IP).  
The {\tt VASP} PBE PP predicts the Ag IP to be 178.9~kcal/mol, while the
first and second IP for Ni are predicted to be 160.6 and 492.9~kcal/mol,
respectively.  Adding the respective $\Delta G_{\rm hyd}$, Eqs.~\ref{ag}
and~\ref{ni} yield $\Delta G$ of +57.5 and~+76.0~kcal/mol, respectively.
These individual half-cell reaction $\Delta G$ have not yet
been referenced to the standard hydrogen potential.
The overall Ag$^+$ + Ni$^{+}$ $\rightarrow$ Ag + Ni$^{2+}$ reaction, however,
does not suffer from surface potential ambiguities.  If we use the IP
predicted using the PBE functional, the $\Delta G$ of this reaction
becomes +18.5~kcal/mol, or +0.80~eV, in water.  We stress that the pertinent
Ag species is the silver atom suspended in water, not bulk silver metal.

PBE predictions for IP are, however, problematic.  While our
pseudopotential PBE method fortuitously predicts an Ag IP
in reasonable agreement with the experimental
value of 174.6~kcal/mol, the most accurate quantum chemistry method
(CCSD(T)) with relativistic corrections in fact underestimates this
value by $\sim 1$~eV.\cite{ag_ip1,ag_ip2}  While the CCSD(T) method 
is accurate for the first IP of Ni,\cite{ni_ip1} our pseudopotential
PBE approach severely overestimated the second Ni ionization
potential measured at 418.7~kcal/mol.\cite{ni_ip2}

A more reasonable approach is to combine experimental IP and AIMD
$\Delta G_{\rm hyd}$.  This yields $\Delta G$=+0.01~eV
for Eq.~\ref{agni}.  The predicted value is significantly more
endothermic than the -0.6~eV cited in the experimental radiolysis
literature.\cite{nenoff,lilie,baxendale}  That -0.6~eV value was
derived by estimating the Ni$^+$ $\Delta G_{\rm hyd}$ using a simple
Pauling ionic radius and a dielectric continuum approximation;\cite{baxendale}
as the authors stressed, ligand field effects, which can be
a fraction of an eV for first row transition metal ions in water,\cite{tm}
were neglected.  AIMD $\Delta G_{\rm hyd}$ calculations, free from
these assumptions, should yield more accurate redox potentials
for metal ions in unstable valence states encountered as transients
in radiolysis experiments.\cite{lilie,baxendale,bax2}

Finally, we note that the Ni$^{2+}$ $\Delta G_{\rm hyd}$ depends on whether
the DFT+U approach is used in calculating $\langle d H(q)/q\rangle_q$ along
the AIMD trajectory.  Setting $U$=4 (6)~eV already decreases the gas
phase Ni$^{2+}$-(H$_2$O)$_6$ cluster binding energy by
$\sim 0.5$~eV (1.0~eV) {\it without} inducing noticeable changes in the
geometry of the complex.  Since the octahedral Ni$^{2+}$ hydration shell
is quite stable in liquid water, a similar change in the aqueous
phase $\Delta G_{\rm hyd}$ is expected if $U$ varies by like amounts.
We have indeed found that using DFT+U ($U$=4~eV) to compute
$\langle d H(q)/q\rangle_q$ decreases the solvation by roughly 12~kcal/mol,
yielding a $\Delta G_{\rm hyd}$ of -353.7$\pm 1.0$~kcal/mol.  With
this DFT+U $\Delta G_{\rm hyd}$, Eq.~\ref{agni} becomes endothermic by
+0.51~eV compared with the +0.01~eV predicted with PBE (i.e., $U$=0~eV).
Whether PBE or DFT+U yields more accurate $\Delta G_{\rm hyd}$
will be assessed in the future by comparison with high level quantum
chemistry, new DFT functionals, \cite{truhlar1}
or gas phase experimental values such as those reported
for monovalent cations and anions.\cite{tiss}

The above analysis suggests that predicting redox potential of half
cell electrochemical reactions of first row transition metal ions like
Ni$^+$ remains a challenge,\cite{zeng,seealso}
and that reported redox values in the radiolysis
literature\cite{lilie,baxendale} may need to be extensively revised.
We stress that our approach, which partitions redox potentials into
hydration free energies and IP, circumvents DFT inaccuracies associated
with IP predictions.
 
\section{Conclusions}
 
We have applied {\it ab initio} molecular dynamics (AIMD) simulations to
compute the absolute hydration free energies of Li$^+$, Cl$^-$,
and Ag$^+$.  While some small contributions from packing (entropy)
effects and simulation cell size dependences for anions still need
to be estimated using classical force field based simulations,
the dominant electrostatic contributions come from
density functional theory (DFT) and rigorous liquid state statistical 
mechanical methods.\cite{hummer_mono,na8,classical,pratt_sur}

To compare with experimental values, care must be taken to account
for surface potential contributions which can be decomposed into
water dipole and quadrupole (``second spherical moment'')
contributions,\cite{pratt_sur,saunders} $q(\phi_d + \phi_q)$.
So far, the water-vapor interface surface potential has not been computed
using AIMD.  Nevertheless, the experimental data tabulated by
Marcus\cite{marcus} and Tissandier {\it et al}.\cite{tiss} can be compared
with AIMD values by adding $q\phi_q$ and subtracting $q\phi_d$ values
estimated using the SPC/E water model, respectively.  In both
cases, we would be comparing with $\Delta G_{\rm hyd}$ values
fitted to the SPC/E water model; but to the extent that the SPC/E
$\phi_d$ is an accurate physical quantity,
comparing AIMD $\Delta G_{\rm hyd}$ with $\Delta G_{\rm Tiss} -
\phi_d({\rm SPC/E})$ (plus a standard state correction C$^{(0)}$)
should be model-independent.  With these caveats, we find that the
AIMD $\Delta G_{\rm hyd}$ for Li$^+$ and Cl$^+$ are within 4.9 (4~\%) and
0.4~kcal/mol (0.5~\%) of Tissandier {\it et al}.'s values adjusted this way.
The deviations from Marcus' values,\cite{marcus} compiled after removing
surface potential and standard state contributions, are larger, probably
due to uncertainties in $\phi_q$ estimates.
The sum of $\Delta G_{\rm hyd}$ for the Li$^+$/Cl$^-$ ion pair, where surface
potential effects cancel, agree with the two sets of experimental values
to within 2.3\% and 2.6\%, respectively.\cite{marcus,tiss}  The Ag$^+$/Cl$^-$
ion pair has a combined $\Delta G_{\rm hyd}$ within 1.2~\% of Marcus' data.

We also compute the change in $\Delta G_{\rm hyd}$ associated
with Ni$^+$ being oxidized to Ni$^{2+}$.  Coupled with the
hydration free energy of Ag$^+$ and experimental ionization potential
values, we arrive at a free energy change of 0.01~eV (PBE) and 0.51~eV
(DFT+U, $U$=4~eV) for the Ag$^+$ + Ni$^+$ $\rightarrow$ Ag (atom) + Ni$^{2+}$
reaction
in water.  Whether PBE or DFT+U yields more accurate $\Delta G_{\rm hyd}$
will be assessed in the future by comparison with high level quantum
chemistry, new DFT functionals, or experimental values.  This calculation
is pertinent to predicting the redox potential of unstable Ni$^+$ ions. 
The Ni$^+$ oxidation potential often cited in the radiolysis experimental
literature actually contains a hydration theoretical free energy estimate
based on the Ni$^+$ Pauling radius, and it does not account for ligand
field effects.\cite{lilie,baxendale}  Our results suggest that such reported
values may need to be re-examined with the more accurate AIMD approach.

Even without more accurate determination of surface potentials, our
formalism can be applied to predict the AIMD $\Delta G_{\rm hyd}$
difference between like-charged ions such as Na$^+$ and K$^+$, which is
relevant to understanding mechanisms of selective ion binding. 
Our work also paves the way for AIMD calculations of the hydration free
energies of more complex ions and of ions at water-material interfaces, inside
carbon nanotubes where material polarizability is significant,\cite{marsman}
and in inhomogeneous aqueous media in general.  Further work on elucidating
the surface potential entirely with AIMD methods, systematic investigation
of the $U$ dependence of hydration free energy
when DFT+U is applied, and comparison with other functionals (e.g.,
BLYP\cite{blyp_ex}) and AIMD packages (e.g., {\tt CPMD}\cite{cpmd}) will
be pursued in the future.

\section*{Acknowledgement}

KL thanks Tina Nenoff and Matt Petersen for useful discussions.
SLR acknowledges funding by
the National Institutes of Health through the NIH Road Map for Medical
Research.
OAvL acknowledges support from SNL Truman Program LDRD project No. 120209.
This work was also supported by the Department of Energy under Contract
DE-AC04-94AL85000, by Sandia's LDRD program.  Sandia is a multiprogram
laboratory operated by Sandia Corporation, a Lockheed Martin Company,
for the U.S.~Department  of Energy.

\newpage

\begin{table}\centering
\begin{tabular}{||c|c|c|c|c|||} \hline
ion    & $N_{\rm water}$  &  $\rho_{\rm water}$ & quadrature  &
$\Delta G_{\rm hyd}$ \\ \hline
Li$^+$   &  32 &  1.00  & 2-pt &  -128.6 \\
Li$^+$   &  32 &  1.00  & 6-pt &  -128.3 \\
Li$^+$   &  64 &  1.00  & 2-pt &  -126.7 \\
Li$^+$   &  32 &  0.97  & 2-pt &  -126.7 \\
Li$^+$   &  32 &  0.97  & 6-pt &  -127.2 \\
Li$^+$   & SPC/E &  1.00  & 6-pt &  -134.9 \\
Li$^+$   & expt$^a$ &  1.00  & NA &  -113.5 \\
Li$^+$   & expt$^{a\dagger}$ &  1.00  & NA &  -137.0 \\
Li$^+$   & expt$^b$ &  1.00  & NA &  -126.5 \\
Li$^+$   & expt$^{b\dagger}$ &  1.00  & NA &  -133.2 \\
\hline
\end{tabular}
\caption[]
{\label{table1} \noindent
Li$^+$ hydration free energies using different computational
protocols.  H$_2$O densities and $\Delta G_{\rm hyd}$
are in units of g/cc and kcal/mol, respectively.
$^a$Ref.~\onlinecite{marcus}; $^b$Ref.~\onlinecite{tiss}.
Experimental values adjusted for surface potentials 
and standard state contributions are marked with a dagger (see text).
}
\end{table}

\newpage

\begin{table}\centering
\begin{tabular}{||c|c|c|c|c|||} \hline
ion    & $N_{\rm water}$  &  $\rho_{\rm water}$ & quadrature  &
$\Delta G_{\rm hyd}$ \\ \hline
Cl$^-$   &  32 &  1.00  & 2-pt &  -79.0 \\
Cl$^-$   &  32 &  1.00  & 6-pt &  -76.6 \\ 
Cl$^-$   &  32$^*$ &  1.00  & 6-pt &  -69.3 \\
Cl$^-$   & 32 SPC/E &  1.00  & 2-pt &  -71.0 \\
Cl$^-$   & 256 SPC/E &  1.00  & 2-pt &  -67.7 \\
Cl$^-$   & expt$^a$ &  1.00  & NA &  -81.2 \\
Cl$^-$   & expt$^{a\dagger}$ &  1.00  & NA &  -65.3 \\
Cl$^-$   & expt$^b$ &  1.00  & NA &  -72.6 \\
Cl$^-$   & expt$^{b\dagger}$ &  1.00  & NA &  -69.7 \\
Li$^+$/Cl$^-$     &  32 &  1.00  & 6-pt & -197.6 \\
Li$^+$/Cl$^-$    & SPC/E &  1.00  & 2-pt & -202.6 \\
Li$^+$/Cl$^-$    & expt$^a$ &  1.00  & NA & -202.3 \\
Li$^+$/Cl$^-$    & expt$^b$ &  1.00  & NA & -202.9 \\
\hline
\end{tabular}
\caption[]
{\label{table3} \noindent
Cl$^-$ hydration free energies.  The asterisk denotes AIMD $\Delta G_{\rm hyd}$
adjusted for finite simulation cell size and packing effects (see 
text).  Also listed are $\Delta G_{\rm hyd}$ for Li$^+$ plus Cl$^-$.  
The SPC/E results for Cl$^-$ and Li$^+$/Cl$^-$ contain the
packing correction.  H$_2$O densities and $\Delta G_{\rm hyd}$
are in units of g/cc and kcal/mol, respectively.
$^a$Ref.~\onlinecite{marcus}; $^b$Ref.~\onlinecite{tiss}.
Experimental values adjusted for surface potentials
are depicted with a dagger; see text for details.
}
\end{table}

\newpage

\begin{table}\centering
\begin{tabular}{||c|c|c|c|c|||} \hline
ion    & $N_{\rm water}$  &  $\rho_{\rm water}$ & quadrature  &
$\Delta G_{\rm hyd}$ \\ \hline
Ag$^+$   & 32 &  0.99  & 2-pt &  -121.3 \\
Ag$^+$   & 32 &  0.99  & 6-pt &  -121.4 \\
Ag$^+$   & 32$^*$ &  0.99  & 6-pt &  -119.8 \\
Ag$^+$   & expt$^a$ &  1.00  & NA &  -102.7 \\
Ag$^+$   & expt$^{a\dagger}$ &  1.00  & NA &  -126.2 \\
Ag$^+$/Cl$^-$    &  32$^*$ &  0.99  & 6-pt & -189.1 \\
Ag$^+$/Cl$^-$    & expt$^a$ &  1.00  & NA & -191.5 \\
Ni+ $\rightarrow$ Ni$^{2+}$ & 32 &  0.99  & 2-pt &  -365.4 \\
Ni+ $\rightarrow$ Ni$^{2+}$ & 32 &  0.99  & 6-pt &  -365.6 \\
Ni+ $\rightarrow$ Ni$^{2+}$$^x$ & 32 &  0.99  & 2-pt &  -354.5 \\
Ni+ $\rightarrow$ Ni$^{2+}$$^x$ & 32 &  0.99  & 6-pt &  -353.7 \\
\hline
\end{tabular}
\caption[]
{\label{table4} \noindent
Ag$^+$ hydration free energies, and Ni$^+$ $\rightarrow$ Ni$^{2+}$
hydration free energy changes.  H$_2$O densities and $\Delta G_{\rm hyd}$
are in units of g/cc and kcal/mol, respectively.  All simulations
are based on the PBE functional, except that the DFT+U formalism
with $U$=4~eV is applied for Ni predictions marked with an ``$x$.''
The asterick denotes $\Delta G_{\rm hyd}$ adjusted for packing effects.
$^a$Ref.~\onlinecite{marcus}.  Experimental values adjusted for surface
potentials are depicted with a dagger; see text for details.
}
\end{table}

\newpage

\newpage

\newpage

\newpage

\newpage

\newpage

\newpage
 
\newpage

\begin{references}

 \bibitem{intro}
R.~T.~Cygan, C.~J.~Brinker, M.~D.~Nyman,
K.~Leung, and S.~B.~Rempe, Mater. Res. Soc. Bull. {\bf 33}, 42 (2007).
 
\bibitem{structure}
Examples include
F.~Brug\'{e}, M.~Bernasconi, and M.~Parrinello, J.~Am.~Soc.~Chem. {\bf 121},
10883 (1999);
S.~B.~Rempe and L.~R.~Pratt, Fluid Phase Equil. {\bf 183-184}, 121 (2001);
L.~M.~Ramaniah, M.~Bernasconi, and M.~Parrinello, J.~Chem.~Phys. {\bf 111},
1587 (1999); 
E.~Schwegler, G.~Galli, and F.~Gygi, Chem.~Phys.~Lett. {\bf 342}, 434 (2001);
I.-F.~Kuo and D.~J.~Tobias, J.~Phys.~Chem.~B {\bf 105}, 5827 (2001);
S.~Raugei and M.~L.~Klein, J.~Am.~Chem.~Soc. {\bf 123}, 9484 (2001);
S.~Raugei and M.~L.~Klein, J.~Chem.~Phys. {\bf 116}, 196 (2002);
K.~Leung and S.~B.~Rempe, J.~Am.~Soc.~Chem. {\bf 126}, 344 (2004);
S.~B.~Rempe, D.~Asthagiri, and L.~R.~Pratt, Phys. Chem. Chem. Phys.
{\bf 6}, 1966 (2004);
S.~Varma and S.~B.~Rempe, Biophys. Chem. {\bf 124}, 192 (2006);
T.~W.~Whitfield, S.~Varma, E.~Harder, G.~Lamoureux, S.~B.~Rempe, and
B.~Roux, J. Chem. Theor. Comput. {\bf 3}, 2068 (2007).
K.~Leung, I.M.B.~Nielsen, and I.~Kurtz, J.~Phys.~Chem.~B {\bf 111}, 4453 (2007).
 
\bibitem{li_rempe}
S.~B.~Rempe, L.~R.~Pratt, G.~Hummer, J.~D.~Kress, R.~L.~Martin, and A.~Redondo,
J.~Am.~Chem.~Soc. {\bf 122}, 966 (2000).

\bibitem{sabo}
D.~Sabo, S.~Varma, M.~G.~Martin, and S.~B.~Rempe, J.~Phys.~Chem.~B
{\bf 112}, 867-876 (2008).

\bibitem{varma}
S.~Varma and S.~B.~Rempe, J.~Am.~Chem.~Soc. {\bf 130}, 15405 (2008).
 
\bibitem{quasichem}
L.~R.~Pratt and R.~A.~LaViolette, Mol. Phys. {\bf 94}, 909 (1998);
L.~R.~Pratt and S.~B.~Rempe, in L.~R.~Pratt and G.~Hummer (Eds.),
{\it Simulation and Theory of Electrostatic Interactions in Solution} (AIP,
New York, 1999), pp.~172-201; T.~L.~Beck, M.~E.~Paulaitis, and L.~R.~Pratt,
{\it The Potential Distribution Theorem: Models of Molecular Solutions}
(Cambridge University Press, New York, 2006);
S. Varma, and S. B. Rempe. Biophysical J. {\bf 93}, 1093 (2007);
S. Varma, D. Sabo,, and S. B. Rempe. J. Molec. Bio. {\bf 376}, 13, (2008).

                                                                                
\bibitem{anatole1}
O.~A.~von Lilienfeld and M.~E.~Tuckerman, J.~Chem.~Theor.~Comput.
{\bf 3}, 1083 (2007); 
O.~A.~von Lilienfeld and M.~E.~Tuckerman, J.~Chem.~Theor.~Comput.
J.~Chem.~Phys. {\bf 125}, 154104 (2006);
O.~A.~von Lilienfeld, R.~D.~Lins, and U.~Rothlisberger, Phys.~Rev.~Lett.
{\bf 95}, 153002 (2005).
 
\bibitem{hummer_mono}
G.~Hummer, L.~R.~Pratt, and A.~E.~Garcia, J.~Phys.~Chem. {\bf 100}, 1206 (1996).

\bibitem{na8}
G.~Hummer, L.~R.~Pratt, and A.~E.~Garcia, J.~Chem.~Phys. {\bf 107}, 9275 (1997).

\bibitem{garde}
S.~Rajamani, T.~Ghosh, S.~Garde, J.~Chem.~Phys. {\bf 120}, 4457 (2004).
 
\bibitem{classical}
G.~Hummer, L.~R.~Pratt, A.~E.~Garcia, B.~J.~Berne, and S.~W.~Rick,
J.~Phys.~Chem. B {\bf 101}, 3017 (1997);
G.~Hummer, L.~R.~Pratt, and A.~E.~Garcia, J.~Phys.~Chem.~A {\bf 102}, 7885
(1998);
H.~S.~Ashbaugh and R.~H.~Wood, J.~Chem.~Phys. {\bf 106}, 8135 (1997);
T.~Darden, D.~Pearlman, and L.~G.~Pedersen, J.~Chem.~Phys. {\bf 109},
10921 (1998);
R.~M.~Lynden-Bell and J.~C.~Rasaiah, J.~Chem.~Phys. {\bf 107}, 1981 (1997);
F.~Figueirido, G.~S.~Del Buono, and R.~M.~Levy, J.~Phys.~Chem. B {\bf 101},
5622 (1997);
P.~H.~H\"{u}nenberger and J.~A.~McCammon, J.~Chem.~Phys. {\bf 110}, 1856
(1999);
A.~Grossfield, P.-Y.~Ren, and J.~W.~Ponder, J.~Am.~Chem.~Soc. {\bf 125}, 15671
(2003);
H.~S.~Ashbaugh and D.~Asthagiri, J.~Chem.~Phys. {\bf 129}, 204501 (2008).

\bibitem{ti0}
J.~G.~Kirkwood, J.~Chem.~Phys {\bf 3}, 300 (1935).

\bibitem{ti}
M.~P.~Allen and D.~J.~Tildesley, {\it Computer Simulation of Liquids}
(Oxford University Press, New York, 1987).
 
\bibitem{fep}
P.~A.~Kollman, Chem.~Rev. {\bf 93}, 2395 (1983).
 

\bibitem{spce}
H.~J.~C.~Berendsen, J.~R.~Gridera, and T.~P.~Straatsma, J.~Phys.~Chem.
{\bf 91}, 6269 (1987).
                                                                                
\bibitem{note1}
Real space truncation of coulomb interactions
lead to other problems.  As our focus is AIMD simulations
based on DFT calculations using periodic boundary conditions, which
almost universally apply Ewald summations, real space truncations
will not be considered further herein.
 
\bibitem{pratt_sur}
L.~R.~Pratt, J.~Phys.~Chem. {\bf 96}, 25 (1992);
M.~A.~Wilson, A.~Pohorille, and L.~R.~Pratt, J.~Chem.~Phys. {\bf 88}, 3281
(1988);
M.~A.~Wilson, A.~Pohorille, and L.~R.~Pratt, J.~Phys.~Chem. {\bf 91}, 4873
(1987);
Y.~Zhou, G.~Stell, and H.~L.~Friedman,  J.~Chem.~Phys. {\bf 89}, 3836 (1988).

\bibitem{tildes}
V.~P.~Sokhan and D.~J.~Tildesley, Mol. ~Phys. {\bf 92}, 625 (1997).
                                                                                
\bibitem{marsman}
K.~Leung and M.~Marsman, J.~Chem.~Phys. {\bf 127}, 154722 (2007).
Our present work closely follows this preceeding work, but uses slightly
different notations.  In particular, instead of ``second spherical moments,'' 
we use the ``quadrupole moments'' more widely used in the liquid state
literature.
 
\bibitem{saunders}
V.~R.~Saunders, C.~Freyria-Fava, R.~Dovesi, L.~Salasco, and C.~Roetti,
Mol.~Phys. {\bf 77}, 629 (1992).

\bibitem{marcus}
Y.~Marcus, Biophys.~Chem. {\bf 51}, 111 (1994), and references therein.
 
\bibitem{tiss}
M.~D.~Tissandier, K.~A.~Cowen, W.~Y.~Feng, E.~Grunlach, M.~H. Cohen, A.~D.
Earhart, J.~V. Coe, and T.~R. Tuttle, J.~Phys~Chem.~A {\bf 102}, 7787 (1998).

\bibitem{pbe}
J.~P.~Perdew, K.~Burke, and M.~Ernzerhof, Phys.~Rev.~Lett. {\bf 77}, 3865 (1996).

\bibitem{dang}
L.~X.~Dang and T.-M.~Chang, J.~Phys.~Chem.~B {\bf 106}, 235 (2002).
 
\bibitem{mundy}
I.~F.~W.~Kuo and C.~J.~Mundy, Science {\bf 303}, 658 (2004).
 
\bibitem{siepmann}
M.~J.~McGrath, J.~I.~Siepmann, I.~F.~W. Kuo, and C.~J.~Mundy,
Mol. Phys. {\bf 104}, 3619 (2006).

\bibitem{barr}
C.~G.~Barraclough, P.~T.~McTigue, and Y.~L.~Ng, J.~Electroanal.~Chem.
{\bf 320}, 9 (1992).

\bibitem{brod}
E.~N.~Brodskaya and V.~V.~Zakharov, J.~Chem.~Phys. {\bf 102}, 4595 (1995).

\bibitem{note9}
Our value for the SPC/E water $\phi_q$ may be slightly different from
values reported in water-vapor interface simulations because of possible
small variations in the water density in interfacial simulation cells.
Note also that Ref.~\onlinecite{tildes} appears to have misquoted
the value of $\phi_d$ for the TIP4P water from Ref.~\onlinecite{brod},
and that the $\phi_d$ for this model reported in Ref.~\onlinecite{pratt_sur}
was computed at T=325~K, not the T=300~K of Ref.~\onlinecite{brod}

 
\bibitem{truhlar}
C.~P.~Kelly, C.~J.~Cramer, and D.~G.~Truhlar, J.~Phys.~Chem.~B
{\bf 110}, 16066 (2006).
 
\bibitem{nenoff}
Z.~Zhang, J.~Huang, D.~T.~Berry, P.~P.~Provencio, and T.~M.~Nenoff,
J.~Phys.~Chem.~C {\bf 113}, 1155 (2009).

\bibitem{nanosyn}
B.~G.~Ershov, E.~Janata, and A.~Henglein, J.~Phys.~Chem. {\bf 98}, 7619 (1994);
J.~Belloni, Catalysis Today {\bf 113}, 141 (2006).
 
\bibitem{asthagiri}
D.~Asthagiri, L.~R.~Pratt, H.~S.~Ashbaugh, J.~Chem.~Phys. {\bf 119}, 2702
(2003).

\bibitem{zeng}
X.~C.~Zeng, H.~Hu, X.~Q.~Hu, A.~J.~Cohen, and W.~T.~Yang, J.~Chem.~Phys.
{\bf 128}, 124510 (2008).

\bibitem{seealso}
J.~VandeVondele, R.~Ayala, M.~Sulpizi, and M.~Sprik,
J.~Electroanal. Chem. {\bf 607}, 113 (2007);
Y.~Tateyama, J.~Blumberger, T.~Ohno, and M.~Sprik, 
J.~Chem.~Phys. {\bf 126}, 204506 (2007).
 
\bibitem{vasp}
G.~Kresse and J.~Furthm\"{u}ller, Phys.~Rev.~B {\bf 54}, 11169 (1996),
Comput.~Mater.~Sci. {\bf 6}, 15 (1996).
 
\bibitem{paw0}
P.~E.~Bl\"{o}chl, Phys.~Rev.~B, {\bf 50}, 17953 (1994).
 
\bibitem{paw}
The {\tt VASP} implementation is discussed in G.~Kresse and D.~Joubert,
Phys.~Rev.~B {\bf 59}, 1758 (1999).
 
\bibitem{water400}
E.~Schwegler, J.~C.~Grossman, F.~Gygi, and G.~Galli, J. Chem. Phys.
{\bf 121}, 5400 (2004); P.~H.-L. Sit and N. Marzari, J. Chem. Phys.
{\bf 122}, 204510 (2005); S.~B.~Rempe, T.~R.~Mattsson, and K.~Leung,
Phys.~Chem. Chem.~Phys. {\bf 10}, 4685 (2008).
 
\bibitem{cpmd}
J.~Hutter et al., CPMD V3.13 Copyright IBM Corp 1990-2008, Copyright
MPI fuer Festkoerperforschung Stuttgart 1997-2001. 

\bibitem{SGPPKrack}
M.~Krack, Theor.~Chem.~Acc. {\bf 114}, 145 (2005).

\bibitem{note3}
We have avoided directly computing the integrand at $q$=0
because the AIMD trajectories may be too short to adequately sample
the small $q$ regions.  $\langle dH(q)/dq\rangle$ at such $q$ values
exhibit larger statistical fluctuations.  To test the extrapolation
to $q$=0, we have conducted classical force field TI simulations with
much longer trajectory lengths but otherwise identical TI protocol
and compared with $\langle dH(q)/dq\rangle$ directly computed at $q$=0.
These simulations indicate that a cubic fit using our set of 6 $q$ values
yields a good approximation to the $q$=0 integrand.

\bibitem{note2}
Recall that this contribution is estimated using
maximally localized Wannier functions to decompose the total
electron density into individual water contributions.\cite{marsman}
As an additional test, we have taken the nuclear configuration of
each of the 32 individual water molecules in an AIMD snapshot,
computed the individual water $\phi_q$ contribution in the absence of other
water molecules, added them, and compared the result with the global
$\phi_q$ correction computed with all 32 H$_2$O simultaneously present
in the same cell.  Even though the individual H$_2$O approach
neglects many-water effects, the two $\phi_q$ contributions computed are within
1~\%, or 1~kcal/mol, of each other.

 
\bibitem{wannier}
N.~Marzari and D.~Vanderbilt, Phys.~Rev.~B {\bf 56}, 12847 (1997).

\bibitem{note}
This behavior is not universal.  Attempting to put a partial (or an entire) 
$2s$ electron on the Li$^+$ pseudopotential to yield Li$^{q+}$ in 
water, as opposed to globally scaling that pseudopotential by the
factor $q$, results in the partial electron leaving the vicinity
of Li$^{q+}$ and becoming solvated as an excess electron in water.
In other words, if we were interested in the Li $\rightarrow$ Li$^+$
half cell reaction, a more complex $\lambda$-paths would have been
needed.  Spontaneous ejection of electrons does not happen with Ag
or Ni$^+$, both of which are less electropositive than Li.  
 
\bibitem{dftu}
V.~I.~Anisimov, J.~Zaanen, and O.~K.~Andersen, Phys.~Rev.~B, 
{\bf 44}, 943 (1991);
A.~I.~Liechtenstein, V.~I.~Anisimov, and J.~Zaanen, Phys.~Rev.~B
{\bf 52}, R5467 (1995).

\bibitem{leung1}
K.~Leung and C.~J.~Medforth, J.~Chem.~Phys. {\bf 126}, 024501 (2007).
 
\bibitem{dftu_aimd1}
P.~H.~L.~Sit, M.~Cococcioni, and N.~Marzari, J.~Electroanal.~Chem.
{\bf 607}, 107 (2007).

\bibitem{b3lyp}
A.~D.~Becke, J.~Chem.~Phys. {\bf 98}, 1372, (1993);
A.~D.~Becke, J.~Chem.~Phys. {\bf 98}, 5648 (1993);
C.~T.~Lee, W.~T.~Yang, and R.~G.~Parr, Phys.~Rev.~B {\bf 37}, 785 (1988).

\bibitem{basis}
The 6-311+G(d,p) basis also yields a PBE binding energy of 15.7~eV.
As discussed in Ref.~\onlinecite{leung1}, 
this basis and the plane-wave/PAW
method used in {\tt VASP} yield results that are in good agreement.


\bibitem{note5}
It may be argued that the trajectory length is short and the sampled
configurations are correlated, which may underestimate the numerical
noise.  Hence we have tested the uncertainty using classical force fields
and much longer trajectories.  With otherwise identical
parameters (32~H$_2$O, 0.1~ps sampling intervals, extrapolation to
$q=0$), a 400~ps SPC/E
trajectory reveals that, on average, a 40~ps segment of the trajectory
exhibits 0.44 and 0.52~kcal/mol standard deviations for the 6- and 2-point
TI scheme, respectively.  Normally a 6-point TI should exhibit far less
noise than a 2-point one; in the present case, the extrapolation
to $q$=0 required for the 6-point trapezoidal rule has introduced
additional uncertainties.  These uncertainties in SPC/E simulations are indeed
comparable to and even smaller than the standard deviations estimated for
the 40~ps AIMD trajectories.


\bibitem{garcia}
G.~Hummer, L.~R.~Pratt, and A.~E.~Garcia, J.~Am.~Chem.~Phys. {\bf 119}, 8523
(1997).

\bibitem{hille}
B~Hille, {\it Ionic Channels of Excitable Membranes} (Sinauer
Associates, Sunderland, MA, 2001).

\bibitem{paliwal}
A.~Paliwal, D.~Asthagiri, L.~R.~Pratt, H.~S.~Ashbaugh, and M.~E.~Paulaitis,
J.~Chem.~Phys. {\bf 124}, 224502 (2006).

\bibitem{hummer_mol_sim}
G.~Hummer, Mol.~Sim. {\bf 28}, 81 (2002).
 
\bibitem{cl_aimd1}
P.~Jungwirth and D.~J.~Tobias, J.~Phys.~Chem.~A {\bf 106}, 379 (2002).

\bibitem{rigid}
K.~Leung and S.~B.~Rempe, Phys.~Chem.~Chem.~Phys. {\bf 8}, 2153 (2006).

\bibitem{note4}
While the isolated Cl$^-$ ion as predicted by PBE may not be stable in
vacuum, within our periodic boundary condition simulations, Cl$^-$
has a well defined total energy and a HOMO-LUMO gap 
over a large range of simulation cell sizes.

\bibitem{sprik_ag}
J.~Blumberger, L.~Bernasconi, I.~Tavernelli, R.~Vuilleumier, and M.~Sprik,
J.~Am.~Soc.~Chem. {\bf 126}, 3928 (2004).

\bibitem{ag_expt1}
J.~Texter, J.~J.~Hastreiter, and J.~L.~Hall, J.~Phys.~Chem. {\bf 87},
4690 (1983).

\bibitem{ag_expt2}
M.~Sandstr\"{o}m, G.~W.~Neilson, G.~Johansson, and T.~Yamaguchi,
J.~Phys.~C: Solid State Phys. {\bf 18}, L1115 (1985).
                                                                                
\bibitem{ag_ff}
V.~Dubois, P.~Archirel, and A. Boutin, J.~Phys.~Chem.~B {\bf 105}, 9363 (2001).

\bibitem{future}
D.~Jiao, K.~Leung, S.~B.~Rempe, and T.~M.~Nenoff, J.~Chem.~Theoret.~Comput.
(submitted).

\bibitem{lilie}
M.~Breitenkamp, A.~Henglein, and J.~Lilie, Ber. Bunsenges. Phys.~Chem.
{\bf 80}, 973 (1976).

\bibitem{baxendale}
J.~H.~Baxendale and R.~S.~Dixon, Z.~Phys.~Chem. (Munich) {\bf 43}, 161
(1964).

\bibitem{tm}
D.~Asthagiri, L.~R.~Pratt, M.~E.~Paulaitis, and S.~B.~Rempe,
J.~Am.~Chem.~Soc. {\bf 126}, 1285 (2004).

\bibitem{bax2}
J.~H.~Baxendale, J.~P.~Keene, and D.~A.~Stott, Chem.~Comm. 
{\bf 20}, 715 (1996).

\bibitem{ag_ip1}
M.~N.~Huda and A.~K.~Ray, Euro.~J.~Phys.~D {\bf 22}, 217 (2003).

\bibitem{ag_ip2}
C.~E. Moore, Natl. Stand.~Ref.~Data.~Ser., Natl.~Bur.~Stand.
(U.S.) 35 (1971).

\bibitem{ni_ip1}
N.~B.~Balabanov and K.~A.~Peterson, J.~Chem.~Phys. {\bf 125}, 074110 (2006).

\bibitem{ni_ip2}
A.~G.~Shenstone, J.~Res.~Natl.~Bur.~Stand. (U.S.) {\bf 74A}, 80 (1970).

\bibitem{truhlar1}
Y.~Zhao and D.~G.~Truhlar, J.~Chem.~Phys. {\bf 125}, 194101 (2006).

\bibitem{blyp_ex}
A.~D.~Becke, Phys.~Rev.~A {\bf 38}, 3098 (1988).

\end{references}
\end{document}